\documentclass[10pt,sigconf,letterpaper, nonacm]{acmart}

\usepackage{amsmath}
\usepackage{amsfonts}
\usepackage{wrapfig}
\usepackage{graphicx}
\usepackage{textcomp}
\usepackage{xcolor}
\usepackage{soul, color}
\usepackage{tabto}
\usepackage{url}
\usepackage{xspace}
\usepackage{booktabs}
\usepackage{multirow}
\usepackage{float}
\usepackage{subcaption}
\usepackage{caption}

\usepackage{algorithm,algorithmicx}
\usepackage{algpseudocode}
\usepackage{multirow}
\usepackage[normalem]{ulem}
\usepackage{wrapfig,lipsum,booktabs}
\usepackage{soul}
\usepackage{xcolor, colortbl}
\usepackage{comment}
\usepackage{float}
 \usepackage{url}

\usepackage[printonlyused]{acronym}
\acrodef{av}[AV]{Autonomous Vehicle}
\acrodefplural{av}[AVs]{Autonomous Vehicles}
\acrodef{etoe}[E2E]{end-to-end}
\acrodef{rtt}[RTT]{Round-Trip-Time}
\acrodef{webrtc}[WebRTC]{Web Real Time Communication}
\acrodef{rtsp}[RTSP]{Real Time Streaming Protocol}
\acrodef{rtp}[RTP]{Real-time Transport Protocol}
\acrodef{pdcp}[PDCP]{Packet Data Convergence Protocol}
\acrodef{rtcp}[RTCP]{Real-time Transport Control Protocol}
\acrodef{phy}[PHY layer]{Physical Layer}
\acrodef{mac}[MAC Layer]{Medium Access Control Layer}
\acrodef{ho}[HO]{Handover}
\acrodefplural{hos}[HOs]{Handovers}
\acrodef{qoe}[QoE]{Quality-of-Experience}\acrodef{tod}[ToD]{\emph{teleoperated driving}}
\acrodef{ran}[RAN]{Radio Access Network}
\acrodef{bler}[BLER]{Block-Level Error Rate}
\acrodefplural{blers}[BLERs]{Block-Level Error Rates}
\acrodef{rb}[RB]{Resource Block}
\acrodefplural{rb}[RBs]{Resource Blocks}
\acrodef{sa}[5G-SA]{5G Standalone}
\acrodef{nsa}[5G-NSA]{5G Non-Standalone}
\acrodef{cqi}[CQI]{Channel Quality Indicator}
\acrodef{odd}[ODD]{\emph{Operational Design Domain}}
\acrodefplural{odd}[ODDs]{Operational Design Domain}
\acrodef{rtk}[RTK]{real-time kinematic}
\acrodef{gps}[GPS/GNSS]{Global Navigation Satellite System}
\acrodef{imu}[IMU]{Inertial Measurement Unit}
\acrodef{cc}[C\&C]{Command and Control}
\acrodef{5gaa}[5GAA]{5G Automotive Association}
\acrodef{us}[U.S.]{United States}
\acrodef{rtsp}[RTSP]{Real-Time Streaming Protocol}
\acrodef{att}[AT]{AT\&T}
\acrodef{tmb}[TM]{T-Mobile}
\acrodef{vz}[VZ]{Verizon}
\acrodef{epc}[EPC]{Evolved Packet Core}
\acrodef{5gc}[5G Core]{5G Core Network}
\acrodef{tdd}[TDD]{Time Division Duplexing}
\acrodef{fdd}[FDD]{Frequency Division Duplexing}
\acrodef{kmph}[km/h]{kilometers per hour}
\acrodef{sae}[SAE]{Society of Automotive Engineers}
\acrodef{3gpp}[3GPP]{3rd Generation Partnership Project}

\acrodef{qos}[QoS]{Quality of Service}

\acrodef{sla}[SLA]{Service Level Agreement}
\acrodef{isp}[ISP]{Internet Service Providers}
\acrodef{bs}[BS]{Base Station}
\acrodef{bsr}[BSR]{Buffer State Report}
\acrodefplural{BS}[BSs]{Base Stations}
\acrodef{qos}[QoS]{Quality of Service}
\acrodef{gnb}[gNodeB]{next Generation Node B}
\acrodefplural{gNB}[gNBs]{next Generation Node Bs}
\acrodef{enb}[eNodeB]{evolved Node B}
\acrodefplural{eNB}[eNBs]{evolved Node Bs}
\acrodef{ue}[UE]{User Equipment}

\acrodef{mcs}[MCS]{Modulation and Coding Scheme}

\acrodef{tti}[TTI]{Transmission Time Interval}
\acrodef{rnti}[RNTI]{Radio Network Temporary Identifier}
\acrodefplural{RNTI}[RNTIs]{Radio Network Temporary Identifiers}
\acrodef{tbs}[TBS]{Tranport Block Size}
\acrodef{rrc}[RRC]{Radio Resource Control}

\acrodef{drb}[DRB]{Data Radio Bearer}
\acrodef{mib}[MIB]{Master Information Block}
\acrodef{cp}[CP]{Control Plane}
\acrodef{dl}[DL]{Downlink}
\acrodef{ul}[UL]{Uplink}
\acrodef{scs}[SCS]{Sub-Carrrier Spacing}

\acrodef{srs}[SRS]{sounding reference signals}
\acrodef{csi}[CSI]{Channel State Information}
\acrodef{crc}[CRC]{cyclic redundancy checks}
\acrodef{roc}[ROC]{rate of change}
\acrodef{tbs}[TBs]{Transport Blocks}
\acrodef{KPIs}[KPIs]{Key Performance Indicators}
\acrodef{ofdm}[OFDM]{Orthogonal frequency-division multiplexing}
\acrodef{harq}[HARQ]{Hybrid Automatic Repeat reQuest}

\acrodef{teleop}[Tele-Op]{Teleoperation}
\acrodef{teleops}[Tele-Ops]{Teleoperations}
\acrodef{ros}[ROS]{Robot Operating System}
\acrodef{quic}[QUIC]{Quick UDP Internet Connections}

\acrodef{fps}[FPS]{Frames Per Second}

\acrodef{urllc}[URLLC]{Ultra-Reliable Low Latency Communication}

\acrodef{pdn}[PDN]{Packet Data Network}
\acrodef{pdu}[PDU]{Packet Data Unit}
\acrodef{mme}[MME]{Mobility Management Entity}
\acrodef{amf}[AMF]{Access and Mobility Management Function}
\acrodef{ausf}[AUSF]{Authentication Server Function}
\acrodef{hss}[HSS]{Home Subscriber Server}
\acrodef{nas}[NAS]{Non-Access Stratum}
\acrodef{smf}[SMF]{Session Management Function}

\acrodef{cups}[CUPS]{Control/User Plane Separation}
\acrodef{rach}[RACH]{Random Access Channel}
\acrodef{sdn}[SDN]{Software Defined Network}
\acrodefplural{sdn}[SDNs]{Software Defined Networks}
\acrodef{rsrp}[RSRP]{Reference Signal Receive Power}
\acrodef{rsrq}[RSRQ]{Reference Signal Receive Quality}
\acrodef{sgw}[SGQ]{Serving Gateway}
\acrodef{gnbcu}[gNodeB-CU]{gNodeB Central Unit}
\acrodef{ntp}[NTP]{Network Time Protocol}
\acrodef{grpc}[gRPC]{Google Remote Procedure Calls}

\acrodef{cav}[CAV]{Connected and Autonomous Vehicle}

\acrodef{psnr}[PSNR]{Peak Signal-to-Noise Ratio}
\acrodef{ssim}[SSIM]{Structural Similarity Index Measure}

\acrodef{obu}[OBU]{on-board unit}

\acrodef{gcc}[GCC]{(Google Congestion Control)}

\acrodef{abr}[ABR]{Adaptive Bitrate}
\acrodef{fec}[FEC]{Forward Error Correction}

\usepackage{afterpage}

\newcommand{\eg}{\textit{e.g.,}\xspace}

\newcommand{\ie}{\textit{i.e.,}\xspace}

\newcommand{\fig}{Fig.~}
\newcommand{\tbl}{Table~}

\newcommand{\cqil}{$CQI_{low}$\xspace}
\newcommand{\cqim}{$CQI_{medium}$\xspace}
\newcommand{\cqih}{$CQI_{high}$\xspace}

\newcommand{\tod}{AV teleoperation\xspace}

\newcommand{\teleop}{teleoperation\xspace}

\newcommand{\teleops}{teleoperations\xspace}

\newcommand{\teleoprt}{teleoperate\xspace}

\newcommand{\simpletitle}[1]{\noindent\textbf{#1}\xspace}

\definecolor{coralpink}{rgb}{0.97, 0.51, 0.47}
\definecolor{spirodiscoball}{rgb}{0.06, 0.75, 0.99}
\definecolor{turquoiseblue}{rgb}{0.0, 1.0, 0.94}
\definecolor{green(pigment)}{rgb}{0.0, 0.65, 0.31}
\definecolor{green(colorwheel)(x11green)}{rgb}{0.0, 1.0, 0.0}
\definecolor{limegreen}{rgb}{0.2, 0.8, 0.2}
\definecolor{fuchsiapink}{rgb}{1.0, 0.47, 1.0}
\definecolor{ceruleanblue}{rgb}{0.16, 0.32, 0.75}
\definecolor{lavenderindigo}{rgb}{0.58, 0.34, 0.92}
\definecolor{mangotango}{rgb}{1.0, 0.51, 0.26}

\newcolumntype{C}[1]{>{\centering\let\newline\\\arraybackslash\hspace{0pt}}m{#1}}

\renewcommand\footnotetextcopyrightpermission[1]{}
\setcopyright{none}

\acmDOI{}
\acmISBN{}
\acmPrice{}

\begin{document}

\title[Teleoperating AVs Over Commercial 5G Networks: Are We There Yet?]{Teleoperating Autonomous Vehicles over Commercial
5G Networks: Are We There Yet?}

\author{Rostand A. K. Fezeu$^{\mathsection}$*, Jason Carpenter$^{\mathsection}$*, Rushikesh Zende$^{\mathsection}$, Sree Ganesh Lalitaditya Divakarla$^{\mathsection}$, Nitin Varyani$^{\dag}$, Faaiq Bilal$^{\mathsection}$, Steven Sleder$^{\mathsection}$, Nanditha Naik$^{\mathsection}$, Duncan Joly$^{\mathsection}$, Eman Ramadan$^{\mathsection}$, Ajay Kumar Gurumadaiah$^{\mathsection}$, Zhi-Li Zhang$^{\mathsection}$}

\affiliation{%
    \institution{$^{\mathsection}$University of Minnesota - Twin Cities  \country{USA} \hspace{2.5em}$^{\dag}$Kalinga Institute of Industrial Technology \country{India}
    }
}

\thanks{
*These authors contributed equally to this paper. \\Corresponding authors: \href{mailto:fezeu001@umn.edu, carpe415@umn.edu?cc=zhzhang@cs.umn.edu,eman@cs.umn.edu}{fezeu001@umn.edu, carpe415@umn.edu}}

\renewcommand{\shortauthors}{Rostand A. K. Fezeu, Jason Carpenter et al.}

\begin{abstract}
    Remote driving, or \emph{teleoperating} \acp{av}, is a key application that emerging 5G networks aim to support. In this paper, we conduct a systematic feasibility study of AV teleoperation over commercial 5G networks from both \emph{cross-layer} and \emph{\ac{etoe}} perspectives. Given the critical importance of \emph{timely delivery of sensor data}, such as camera and LiDAR data, for AV teleoperation, we focus in particular on the performance of uplink sensor data delivery. We analyze the impacts of \ac{phy} 5G radio network factors, including channel conditions, radio resource allocation, and \acp{hos}, on \ac{etoe} latency performance. We also examine the impacts of 5G networks on the performance of upper-layer protocols and \ac{etoe} application \ac{qoe} adaptation mechanisms used for real-time sensor data delivery, such as \ac{rtsp} and \ac{webrtc}. 
Our study reveals the challenges posed by today's 5G networks and the limitations of existing sensor data streaming mechanisms. The insights gained will help inform the co-design of future-generation wireless networks, edge cloud systems, and applications to overcome the low-latency barriers in AV teleoperation.

\end{abstract}


\maketitle

\section{Introduction}
\label{s:intro}

Since the DARPA Grand Challenge in 2005~\cite{DARPA-AV-challeng}, tremendous progress has been made in the development of \acp{av}. Pilot ``robotaxi'' services are now available in several major cities in the \ac{us}~\cite{SF-robotaxi-NPR,robotaxi-Austin,robotaxi-Phoenix}. Today's \acp{av} can, at best, be rated as \ac{sae} Level-4~\cite{2016-01-0128,sae2014automated,sae2018surface,sae2018taxonomy}: Namely, such \ac{av} is designed with a specific set of conditions, referred to as its \ac{odd}, outside which it must come to a safe stop. Despite significant advances in AI/ML, fully autonomous driving (Level-5) still has a long way to go.
Safety concerns and other issues plaguing robotaxi trials~\cite{Verge-SF-robotaxi-1,Verge-SF-robotaxi-2,Cruise-SF-accident-ceo-resign,Cruise-SF-accident} highlight the challenges posed by complex real-world environments. 

To partly circumvent
these challenges, remote driving -- or \ac{tod} in the \ac{3gpp} parlance -- has been proposed as
an alternative or complementary approach~\cite{5GAA-ToD-system-requirements-analysis,Tao-Zhang-IEEE-IoT,IEEE-IoT-11}, where a remote human
operator takes over the control of \ac{av} when needed. For example, before the \ac{av} is about
to encounter a situation outside its \acsu{odd} and has to stop~\cite{how-self-driving-cars-get-help}. The potential of \ac{tod} is inspired by the promise of 5G, and is considered one of its key use cases by 3GPP and \ac{5gaa}~\cite{5GAA-ToD-system-requirements-analysis}. \ac{tod} has
been tested in (mostly ideal and) restricted environments~\cite{Cell-Fusion-Ref-6,harris2018ces,samsung2019first,European-study-2,ni2023cellfusion}; several
start-up companies are promoting remote driving for certain use
cases \cite{German-Vay,Las-Vegas-Halo,guident,IEET-IoT-12}. Therefore we ask, 
\emph{Can today's commercial 5G meet the requirements of \ac{tod} in real-world environments?} This question is the main goal of this paper.

To this end, we contribute to the understanding of \ac{tod} and carry out a systematic feasibility study of \ac{av} \teleops over commercial 5G networks in real-world urban environments in the city of Minneapolis in the \ac{us} We note that \ac{av} \ac{tod} requires \emph{timely delivery} of i) \ac{ul} on-board sensor data like camera/video and LiDAR feeds from a host \ac{av} to a remote vehicle control station operated by a human to provide \emph{(real-time) situation awareness}; and ii) \ac{dl} \ac{cc} data from the remote human operator to the host \ac{av} for vehicle control remotely. Per \cite{5GAA-ToD-system-requirements-analysis} (see~\S\ref{ss:av-req}), the \ac{etoe} \ac{ul} and \ac{dl} latency within 100~ms and 20~ms respectively are considered ideal. In terms of data rates, clearly \ac{ul} sensor data -- video and especially LiDAR -- require significantly high bandwidth, whereas \ac{dl} delivery of \ac{cc} data requires little bandwidth (\S\ref{ss:av-req}). Unfortunately, today's commercial 5G networks are designed, configured, and optimized primarily for mobile Internet access. This creates \emph{\ac{dl}/\ac{ul} asymmetry in bandwidth and latency} (see~\S\ref{ss:5g-req}). Furthermore, 
\emph{high mobility} of \acp{av} induces highly fluctuating radio channel conditions and frequent \acp{hos}, both of which can increase \acp{bler} at the \ac{ran} and require retransmissions, thereby incurring higher latency.

We take a \emph{cross-layer} approach to investigate the fundamental impacts of 5G networks on the \ac{av} teleoperation. Given the stringent performance requirements of teleoperation, our aim is to characterize the \ac{etoe} performance for sensor data -- video and LiDAR -- streamed over a commercial 5G network to an edge-cloud server in a remote teleoperation station in the same geographical region as the \ac{av}. We introduce \ac{qoe} metrics at the \emph{per-frame} level -- as frames are the basic units for video display and camera/LiDAR data processing (\S\ref{ss:qoe}). We investigate the effectiveness of video compression (\S\ref{ss:single_camera}) and bitrate adaptation (\S\ref{s:cc-adaptation}) in reducing the one-way delay when streaming video (and LiDAR) data, and quantify timing of critical \ac{cc} including steering, acceleration, and braking over commercial 5G networks (\S\ref{ss:cc}). We extract 5G \ac{ran} parameters to understand the impact of 5G \ac{phy} factors -- such as channel conditions, \acp{bler}, \acp{hos}, and \acp{rb} allocation -- on per-frame \ac{etoe} latency, with a focus on \emph{tail latency} (\S\ref{s:5gRAN}). Additionally, we evaluate how 5G affects upper-layer protocols and \ac{etoe} application layer \ac{qoe} adaptation mechanisms used in real-time sensor data delivery protocols like \ac{rtsp}~\cite{tool-rtsp} and \ac{webrtc}\cite{webrtc}. From our insights, we further explore challenges in teleoperating multiple \ac{av}s over 5G and examine potential benefits of using multiple 5G operators to mitigate tail latency degradation and optimize network resource utilization.

\noindent
The key findings of our paper are summarized as follows:

\noindent
$\bullet$ We define \emph{per-frame} level \ac{qoe} metrics to characterize sensor data delivery \ac{etoe} latency performance and visual quality. In particular, we introduce \emph{perceptual quality deviation} to capture the latency-visual-quality interplay (\S\ref{ss:qoe}). Using \ac{rtsp}
(commonly used in many existing \ac{av} \teleop platforms~\cite{telecarla,teleop-ftm-tod}) 
and \ac{webrtc} (a common low latency video conferencing streaming system, and used by industry~\cite{guident})
as the baselines~\S\ref{ss:single_camera}, we examine the effectiveness of video data compression and LiDAR data voxel-based downsampling techniques in attaining \ac{etoe} low-latency performance and quantify the impact of latency on \emph{perceptual quality} performance. We find that while it is feasible to stream a single camera (as is the case in several field tests~\cite{European-study-2,ni2023cellfusion}) to \teleoprt an \ac{av} in most scenarios, the poor \emph{tail latency performance} still raises safety concerns, as a safety-critical event can occur during such periods of poor performance.

\noindent
$\bullet$ Situational awareness for teleoperation requires streaming multiple camera feeds and LiDAR data. We find that streaming these sensors simultaneously would likely strain today's 5G networks, even with the \ac{sa} architecture. Without aggressive compression, streaming LiDAR over today's 5G networks is nearly impossible (\S\ref{ss:lidar}). We also find that all existing data streaming mechanisms suffer from the \emph{cumulative latency effect} (\S\ref{ss:single_camera}), which causes later frames to be more likely to suffer from deadline violations. 

\noindent
$\bullet$ For \ac{dl} \ac{cc}, the latency experienced is roughly half of the \ac{ul} latency and does not face throughput bottlenecks due to the smaller sizes of the commands. This echoes existing work and strengthens the feasibility of sending control commands with reasonable time and reliability (\S\ref{ss:cc}).

\noindent
$\bullet$ 5G \ac{phy} dynamics significantly affect sensor data delivery for teleoperation. For example, on the one hand, when the channel conditions, as characterized by \ac{cqi}, go from ``good'' to ``poor'', the \ac{etoe} frame delay increases by about 48\%. In contrast, when fewer retransmissions occur on the \ac{phy} -- quantified by a shift in \acp{bler} from +10\% down to 0-5\%, the frame delay drops by approximately 35.8\%. On the other hand, \acp{hos} pose an even greater challenge; we find that unnecessary ping-pong \acp{hos} can occur within a short time window (15 seconds in our analysis) while driving in a loop and making turns. As a result, the sensor data frame delay increases by up to 85\% (\S\ref{s:5gRAN}).

\noindent
$\bullet$ Our results show that \ac{webrtc}, Google's live video streaming system, succeeds in delivering low latency video frames (compared to \ac{rtsp}), but fails to quickly respond to changing 5G network conditions resulting in overall reduced and struggling performance despite the \ac{phy} awareness and speediness of recovery. \ac{rtsp} lacks any of these mechanisms and therefore struggles significantly in terms of per-frame delay impact. This manifests as a spike in per-frame latency, which could have been avoided with the knowledge of the lower-layer information. As we demonstrate in \S\ref{s:cc-adaptation}, the lower layer was aware of a problem far sooner than the application-level congestion and \ac{qoe} triggers.

\noindent
$\bullet$ 
Based on the above findings, we briefly explore the potential benefits of additional end-system mechanisms, such as selective frame dropping/frame rate adaptation, as well as leveraging multiple 5G operators to improve tail latency performance. We also consider the additional challenges posed by multiple AVs competing for radio resources (\S\ref{s:e2eApproaches}). \\

\simpletitle{Contributions.} We contribute to advancing the understanding of teleoperated driving over commercial 5G networks and conduct -- to the best of our knowledge -- the first feasibility study of \ac{av} teleoperation over \emph{operational} live commercial 5G networks in a \emph{real-world} urban environment from \emph{cross-layer} and \emph{\ac{etoe}} perspectives, elucidating in particular the impacts of 5G networks on \ac{ul} vehicle sensor data delivery. Our study reveals the challenges posed by 5G networks as well as the limitations of existing sensor data streaming mechanisms. Through exploration of additional end-system mechanism designs, we show that while these mechanisms can improve the tail latency performance, they cannot fundamentally address the challenges posed by 5G networks.

\begin{figure*}[t!]
    \begin{minipage}[c]{0.72\textwidth}
        \centering
        \includegraphics[scale=0.45, keepaspectratio]{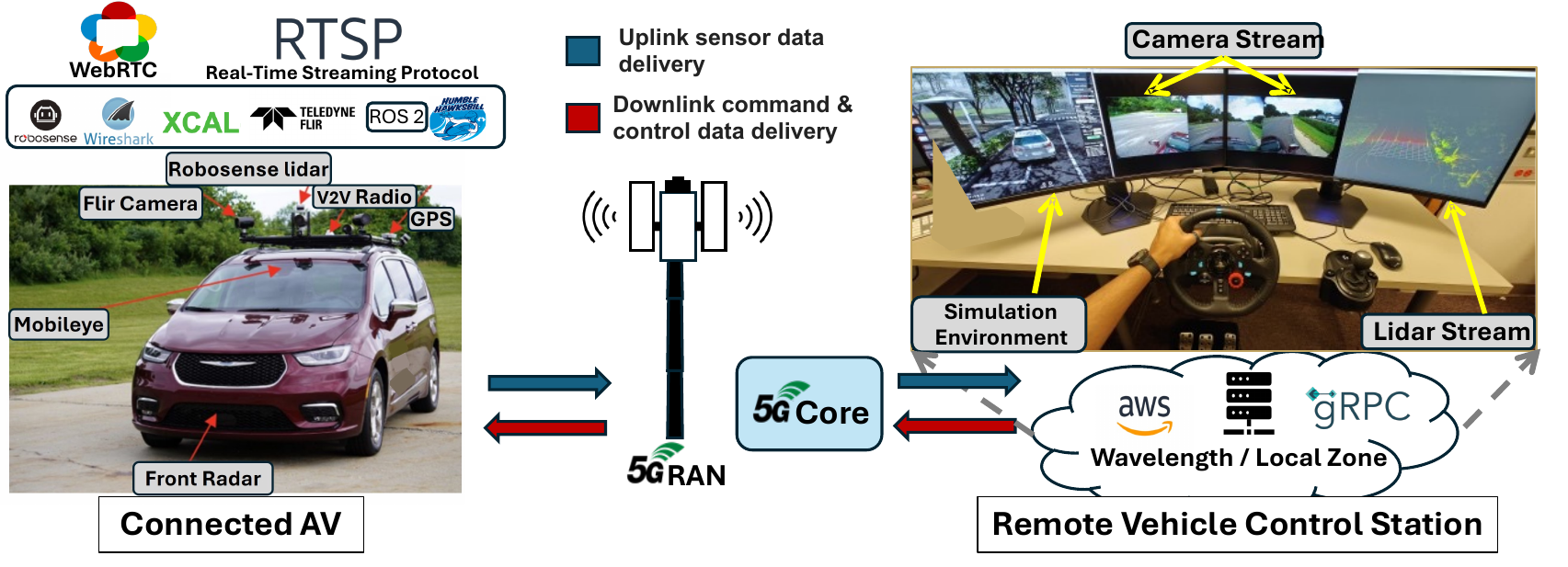}

         % \vspace{-3ex}
         \caption{Experimental Setup, Tools, \& Streaming System.}
         \label{fig:setup}

    \end{minipage}
    \hfill
    \begin{minipage}[c]{0.27\textwidth}%
        \captionof{table}{\ac{av} Teleoperation Latency Requirements}
        \vspace{-2.5ex}
      \label{tab:teleop-req}
      \centering
      \setlength{\tabcolsep}{2pt}
      \setlength{\extrarowheight}{2pt}
      \resizebox{0.85\columnwidth}{!}{
      \centering
      \footnotesize
      \begin{tabular}{c | c}
      \toprule 
      \midrule[.1em]
      \textbf{\begin{tabular}[c]{@{}c@{}}\large Application \\ \large Level\end{tabular}} & \textbf{\begin{tabular}[c]{@{}c@{}}\large 5G Network \\ \large Level\end{tabular}} 
      \\ \hline
       \cellcolor{teal!20} \begin{tabular}[l]{@{}l@{}}\large UL: 100~ms\end{tabular} & 
       \cellcolor{teal!20} \begin{tabular}[l]{@{}c@{}}\large UL: 40 - 45~ms\end{tabular} 
      \\ \hline
      \cellcolor{orange!12} \begin{tabular}[l]{@{}l@{}}\large \large DL: 20~ms\end{tabular} & 
      \cellcolor{orange!12} \begin{tabular}[l]{@{}c@{}}\large \large DL: 15~ms\end{tabular} 
      \\
      \bottomrule
  \end{tabular}
  }
        
        \bigbreak
        \vspace{-2ex}
        
     \centering
    \captionof{table}{Sensor Data Throughput Requirements (in Mbps).}
    \vspace{-2.5ex}
    \label{tab:av-req}

    \setlength{\tabcolsep}{3pt}
    \setlength{\extrarowheight}{2pt}
    \resizebox{0.99\columnwidth}{!}{
    \centering
    \footnotesize
    \begin{tabular}{cc| cc| c}
    \toprule 
    \midrule[.1em]
    
    \multicolumn{2}{c|}{\cellcolor{purple!20} \textbf{\begin{tabular}[c]{@{}c@{}}\large Camera \large (UL)\end{tabular}}} &
    \multicolumn{2}{c}{\textbf{\cellcolor{yellow!40} \begin{tabular}[c]{@{}c@{}}\large LiDAR \large (UL)\end{tabular}}} &
     \cellcolor{blue!20} {\textbf{\begin{tabular}[c]{@{}c@{}}\large C\&C \large (DL)\end{tabular}}}
    \\ \hline
   \multicolumn{1}{c|}{\cellcolor{purple!20} \begin{tabular}[c]{@{}c@{}}\large 1-\\ \large Steam\end{tabular}} &
   \cellcolor{purple!20} \begin{tabular}[c]{@{}c@{}}\large 4-\\ \large Streams\end{tabular} &
   \multicolumn{1}{c|}{\cellcolor{yellow!40} \begin{tabular}[c]{@{}c@{}}\large 64-\\ \large Beams\end{tabular}} &
   \cellcolor{yellow!40} \begin{tabular}[c|]{@{}c@{}}\large 128-\\ \large Beams\end{tabular} &
   \cellcolor{blue!20} \begin{tabular}[c]{@{}c@{}} \\ \end{tabular}
   \\ 
    \cline{2-5} \hline
   \multicolumn{1}{c|}{\cellcolor{purple!20} \large 8} & 
   \large \cellcolor{purple!20} 32 &
   \multicolumn{1}{c|}{\cellcolor{yellow!40} \large 277} &
   \cellcolor{yellow!40} \large {307} &
   \cellcolor{blue!20} \large 0.3 
   \\ 
    \bottomrule
    \end{tabular}
    }

    \end{minipage}
     \vspace{-1ex}
\end{figure*}

\section{Background and Motivation}
\label{s:capabilities}

\noindent

\fig~\ref{fig:setup} shows our testbed, a typical setup -- an \ac{av} equipped with several sensors, sending data via 5G in the \ac{ul} to a remote vehicle control station. A remote teleoperator then issues a maneuver, trajectory \ac{cc} in the \ac{dl} to the vehicle.  The delivery of sensor and \ac{cc} data via the 5G networks has strict latency requirements, which if not met, can be dangerous and potentially even catastrophic, especially in unexpected time-critical situations facing the vehicle. In this section, we discuss the application-level requirements for teleoperation and the required throughput to stream vehicle sensor data. We then provide some background on existing commercial 5G networks and motivate our feasibility study. 

\begin{figure*}[t]
\begin{minipage}{0.35\textwidth}%
\centering
   % \vspace{1ex}
  \includegraphics[scale=0.22,keepaspectratio]{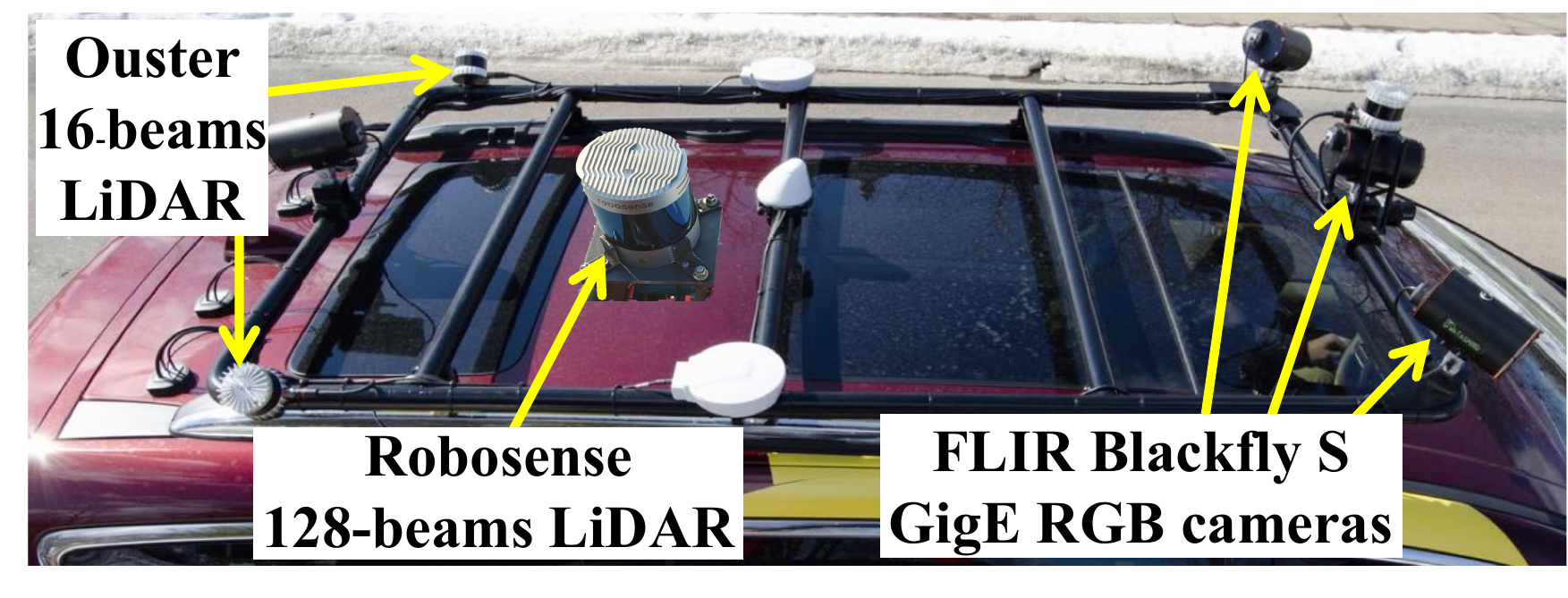}
 % \vspace{-5ex}
  \caption{\ac{av} top-view with Sensors.}
  \label{fig:mn-cav}
    
\end{minipage}
\hspace{0.1cm}
\begin{minipage}{0.32\textwidth}
    \centering
    \includegraphics[width=0.99\textwidth, keepaspectratio]{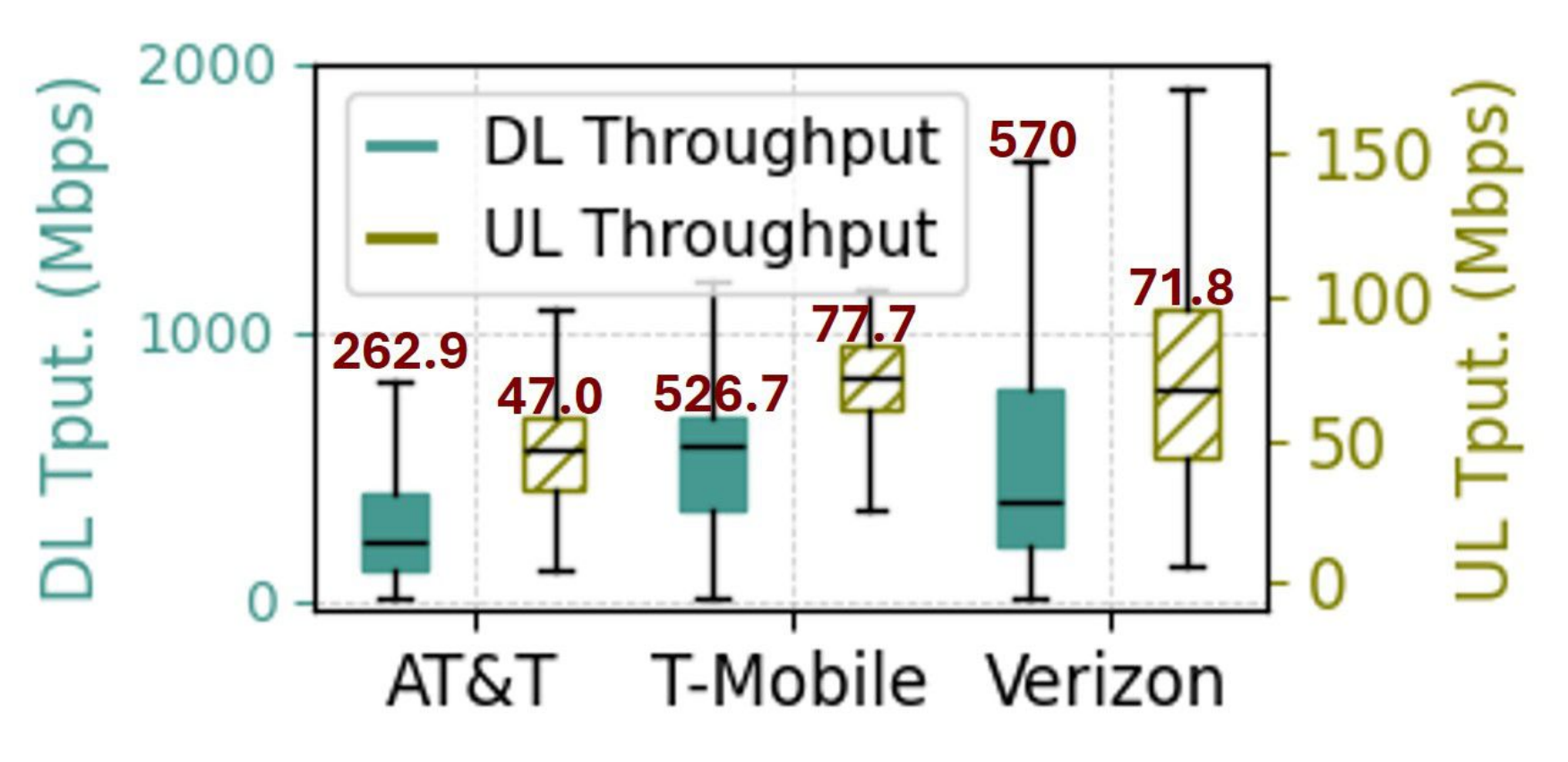}
     \vspace{-0.28in}
    \caption{PHY \ac{dl} \& \ac{ul} Throughput.}
    \label{fig:dl-ul-tput}
    
\end{minipage}
\hspace{0.1cm}
\begin{minipage}{0.29\textwidth}
    \centering
   % \vspace{1ex}
    \includegraphics[width=0.97\textwidth,keepaspectratio]{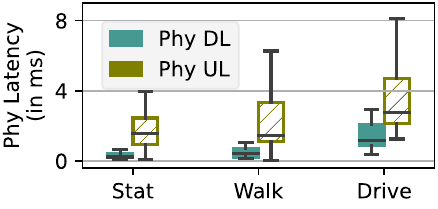}

    % \vspace{-1ex}
    \caption{PHY \ac{dl} \& \ac{ul} Latency.}
    \label{fig:phy-delay}
    
\end{minipage}
 %\vspace{-3ex}
\end{figure*}

\simpletitle{Latency Requirements. }\tbl~\ref{tab:teleop-req} summarizes the latency requirements as specified by \ac{5gaa}~\cite{5GAA}. These requirements are derived from system analysis, simulation/emulation studies, and field experiments (see~\cite{5GAA-ToD-system-requirements-analysis,5GAA-C-V2X-use-case-requirement-vol-2,5GCroCo,RoboAuto,Bennett2020}). Clearly, the \ac{dl} \ac{cc} data has a more stringent delay requirement of 20~ms and 15~ms for the application-level and 5G network-level respectively, but consumes far less bandwidth (\ie 0.3~Mbps). In the \ac{ul}, the application-level latency requirement is 100~ms while the 5G network-level latency requirement is 40--45~ms. These requirements are also in line with the observations gained from emulation-based human subject studies in~\cite{Raj-tele-op-paper, 5GCroCo,RoboAuto,Bennett2020} -- in which, a human driver can generally steer the vehicle remotely when the application-level \ac{ul} latency is under 100~ms, but performance degrades quickly beyond that, becoming nearly impossible above 500~ms.

\vspace{-0.5em}

\subsection{\ac{tod} Performance Requirements}
\label{ss:av-req}
\simpletitle{AV Sensor Data Rate Requirements. }We use our \ac{av}'s top view, shown in \fig~\ref{fig:mn-cav}, as an example of some sensors in a typical \ac{av}. The \ac{gps}, enhanced by \ac{rtk} positioning and onboard \ac{imu} and odometer readings, provides accurate (centimeter-level) data on location, speed, orientation, and angular rate—essential for trajectory tracking and driving control. The GigE RGB cameras capture wide road views for situational awareness and object detection, while a thermal camera aids nighttime detection by showing heat signatures. LiDAR uses light reflections to measure distances and create high-resolution 3D point clouds. When augmented with Camera data feeds, the depth information can be associated with each object, giving the remote operator a sense of how far each object is. Radar, though less precise, detects objects by measuring distance, azimuth, and velocity using radio waves, offering a longer range than LiDAR. Refer to~\tbl~\ref{table:raw_throughput} in Appendix~\ref{aa:method} for a detailed discussion of all our \ac{av} sensors. According to~\cite{lidar_video_size} and our data collection, LiDAR and camera data constitute 78.3\% and 18\% of all sensor data streamed by a typical \ac{av}, while all other data types constitute $<$~1\%. Thus, our analysis in this paper mainly focuses on camera and LiDAR data with throughput requirements of 8~Mbps for a single camera, 277~Mbps and 307~Mbps for a 64- and 128-beams LiDAR data, respectively (see~\tbl~\ref{tab:av-req}).

\vspace{-1ex}

\subsection{5G Networks Today: PHY Performance}
\label{ss:5g-req}

\noindent
Commercial 5G networks are widely deployed worldwide, and a number of measurement studies have been published characterizing their performance (see \eg~\cite{Ross-5G-mid-band-arxiv, ross-sigcomm, hassan2022vivisecting, measurment-5g-ul-1}, and~\S\ref{s:related} for more discussion). In Minneapolis, where we conduct our \ac{av} feasibility study, all three major \ac{us} operators, \ac{att}, \ac{tmb}, and \ac{vz}, are deployed. At the time of our study, \ac{att} and \ac{vz} deployed their 5G service using the \ac{nsa} mode -- which depends on 4G \ac{epc}. \ac{att} and \ac{vz} utilize primary 5G mid-band (more specifically, the C-band) and mmWave channels, respectively. In contrast, since spring 2023 \ac{tmb} deployed its 5G service in the \ac{sa} mode -- which depends on \ac{5gc}. \ac{tmb} utilizes multiple mid-band channels (in band n25, n41) as well as a few low-band channels (in band n71). All mid-band channels (with the exception of \ac{tmb} n25 band) and mmWave high-band channels use the \ac{tdd} mode for their 5G services; whereas the \ac{tmb} n25 and low-band channels use the \ac{fdd}.

\simpletitle{``Best'' Achievable 5G Throughput.} To gauge the feasibility of teleoperation in commercial 5G networks today, we measure the ``best'' (max) 5G throughput today by conducting a series of repeated bulk data transfers using iPerf3~\cite{tool-iperf3} (see~\S\ref{s:method} for our detailed measurement campaign). These experiments are conducted under mobility, where the \ac{av} is traveling at 16 -- 65 \ac{kmph} with stop-and-go at intersections. \fig~\ref{fig:dl-ul-tput} shows the \ac{dl} and \ac{ul} \ac{phy} layer throughput performance. We see \ac{dl} throughput is significantly higher than the \ac{ul} throughput for all operators (notice the different scale of y-axis for \ac{dl} and \ac{ul}). Notably, while the average \ac{dl} \ac{phy} throughput of \ac{att}, \ac{tmb}, and \ac{vz} are {262.9}~Mbps, {526.7}~Mbps, and {570}~Mbps respectively, their \ac{ul} PHY throughput are {47}~Mbps, {77.7}~Mbps, and {71.8}~Mbps respectively. Among all three operators, the best (\ie \emph{peak}) \ac{dl} and \ac{ul} \ac{phy} throughput performance of \ac{vz} ({1639}~Mbps and {174}~Mbps) is due to its mmWave radio band which is highly sensitive to obstruction~\cite{narayanan2020firstlook}. Such significant \emph{\ac{dl}/\ac{ul}} asymmetry in throughput is, in a sense, by design, largely coming from the \ac{tdd} frame structure~\cite{Ross-5G-mid-band-arxiv}. This is mainly because today's commercial 5G networks (as in 4G LTE networks) are ``optimized'' for mobile Internet access, where the majority of applications are \ac{dl}-centric. This is problematic for \ac{ul}-centric applications, such as teleoperation.

\simpletitle{Achievable 5G Latency. }Inspired by~\cite{rossPam, Eiman_2020, ross-sigcomm}, we quantify the 5G PHY layer (``over-the-air'' data) latency defined as; the \textit{PHY \ac{dl} $+$ \ac{ul} latency}. In \fig~\ref{fig:phy-delay}, we show the PHY \ac{ul} and \ac{dl} latency for \ac{tmb} when stationary, walking, and driving. Again, we see a clear \emph{\ac{dl}/\ac{ul} asymmetry in latency}. Mobility further worsens latencies. These \emph{\ac{dl}/\ac{ul} asymmetry} and mobility undoubtedly pose significant challenges for teleoperation today. At the time of our study, \ac{tmb} achieves the ``smallest'' average \ac{dl} and \ac{ul} latency probably due to its \ac{sa} architecture~\cite{ye2023closer}. Thus, unless otherwise mentioned, our analysis mainly focuses on \ac{tmb}. 

\simpletitle{\textbf{\textit{Feasibility of Teleoperation. }}} In total, based on these \ac{av} sensor data requirements and ``peak'' achievable 5G throughput today, one could paint a picture of the feasibility (or infeasibility) of teleoperation over commercial 5G today. Using \ac{tmb} as an example which boasts an \ac{ul} throughput of {77.7}~Mbps, streaming 4 camera streams plus one 64 beams LiDAR with a {309}~Mbps total \ac{ul} throughput requirements (see~\tbl~\ref{tab:av-req}), would be unrealistic within the stringent latency requirements (see~\tbl~\ref{tab:teleop-req}). Furthermore, the actual performance in the wild would vary dramatically under the weight of all sensors streaming, highly diverse channel conditions, network resource competition, and more.

\section{Related Works}
\label{s:related}

\simpletitle{5G Measurements. }There is a plethora of in the wild measurement studies in China~\cite{xu2020understanding}, the \ac{us}~\cite{ye2023closer, measurement-5g-general-2, hassan2022vivisecting, measurment-5g-ul-1, narayanan2020firstlook,narayanan2020lumos5g,ramadan2021videostreaming,5g-meas-chicago-miami,narayanan2022comparative,rossPam,liu2023unrealizedpotentials, ghoshal-imc23, carpenter2023multi, rochman2023comprehensive, weiye2024ca}, and Europe~\cite{roaming-paper, 5g-mmsys-europe, fiandrino2022uncovering, ross-sigcomm, Ross-5G-mid-band-arxiv}. All characterize the performance of commercial 5G networks from the user's perspective across different layers in various scenarios including mobility, when roaming abroad, with high traffic load, and when streaming. 

Most of these works have considered \ac{dl}-centric applications. Relevant to our work, are studies that aim to understand \ac{ul}-centric applications. For instance, \cite{ghoshal-imc23, measurment-5g-ul-1} (and others) studied large data upload, camera sensor delivery, and other killer-apps like AR/VR and \ac{cav}. All these studies revealed important insights in terms of camera sensor data streaming over commercial 5G networks, analyzing several factors like channel conditions, handovers, edge server placement for \ac{ul}-centric applications. However, these studies; i)~did not consider a realistic testbed equipped with several sensors including LiDAR data streaming, down-sampling and voxel-based compression techniques for real-sensor data delivery, and ii)~to a great extent failed to offer an in-depth discussion of the intricate interplay of streaming multiple sensor data over commercial 5G -- for instance, highlighting the challenges that several factors like \acp{bler}, handovers, and channel conditions while driving and taking sharp turns ($\leq$90 turns) presents for \tod today. 

\simpletitle{AV Teleoperation. }Several studies~\cite{ni2023cellfusion, carpenter2023multi, teleop-4g-5g-1,European-study-2,teleop-ftm-tod, teleop-backup-remote-op} have also studied \tod over cellular networks. Among them, \cite{carpenter2023multi, teleop-backup-remote-op} are the most relevant. In our preliminary work~\cite{carpenter2023multi}, we conducted an in-depth measurement study collecting multi-modal vehicle sensor data, including video, LiDAR, and 5G network metrics over thousands of kilometers using an emulation testbed. This analysis uncovered challenges in supporting multi-modal streams due to the variable 5G throughput. \cite{teleop-backup-remote-op} used \ac{webrtc} and OMNeT++ 5G network simulator to improve \ac{qoe} for a remote-controlled ferry while sailing. 

Although these past works advance the understanding of \tod over 5G, i)~they are still in the simulation/emulation phase instead of real-world sensor data streaming, and ii)~there is a noticeable absence of cross-layer analysis and the connection between network performance and \tod \ac{qoe}. An understanding of the underlying causes of \tod performance to establish a causal relationship between network performance and \ac{qoe} has yet to be fully explored. Our paper contributes to the understanding of \tod over live 5G networks. We provide valuable insights to bridge the existing gap in this area.

\section{Experimental Setup, Tools, and Streaming QoE Metrics}
\label{s:method}

\subsection{Experimental Setup \& Data Collection}
\label{ss:exp-setup}

\simpletitle{Testbed and Measurement Platforms. }In~\fig~\ref{fig:setup}, we illustrate our experimental testbed. The \ac{av} is equipped with an on-board computer to transmit sensor data. For remote vehicle control stations, we rely on an Amazon AWS Cloud~\cite{AWS} server deployed in an AWS Local Zone in the same geographical area as our \ac{av}. The AWS Local Zone is the second-nearest ``edge'' server, with a \ac{rtt} of 36.60~ms±4.58~ms (using \ac{tmb}) from our \ac{av}. 

\simpletitle{Streaming System. }We perform a series of live video, LiDAR, and \ac{cc} data streaming over 5G. i)~In the \ac{ul}, we stream camera data (at 30~\ac{fps} with a single camera feed and a multi-camera merged option) using  \ac{rtsp} and \ac{webrtc}, both use \ac{rtp} over UDP. We chose \ac{webrtc} because, unlike \ac{rtsp} which does not perform any bitrate adaptation, \ac{webrtc} is designed to achieve lower latency by incorporating congestion control detection and bitrate adaptation mechanisms, and currently being used by industry~\cite{guident} for commercial teleoperation. For LiDAR data streaming, we rely on the \ac{ros}. The data generated by the LiDAR is encapsulated as \ac{ros} messages for streaming. ii)~For \ac{dl}, we send \ac{cc} data using the \ac{grpc} framework\cite{grpc}. iii)~To study the effect of data compression on teleoperation, we utilize various video compression algorithms on both \ac{rtsp} and \ac{webrtc}, and employ voxel-based downsampling of the LiDAR data.

\simpletitle{Measurement Tools. }Since our goal is to perform a cross-layer analysis of \tod, we need to collect data at the application layer and the 5G network. i)~We rely on Accuver XCAL~\cite{xcal} -- a commercial grade tool which collects detailed 5G \ac{ran} protocol stack information. ii)~At the application layer, using \ac{rtsp} and \ac{webrtc}, we implement logging on both the \ac{av} and teleoperator control server-side. Furthermore, we rely on Wireshark~\cite{tool-wireshark} data, captured on both the \ac{av} and server-side to compute the streaming \ac{qoe} and \ac{cc} delays  metrics (See ~\S\ref{ss:qoe} and~\S\ref{ss:cc} respectively). iii)~We synchronize the clocks of the \ac{av} on-board computer and the AWS server with \ac{ntp} to ensure correctness of clocks and timing information.

\simpletitle{Data Collection Approach. }Conducting experiments in the wild using our research \ac{av} is not only very (cost and labor) expensive, but is also very challenging. For this reason, we first conduct a 1748~Km (1086-miles) cross-state driving measurement campaign that spans 5 days, in which our \ac{av} is driven by a human driver and we configured the on-board computer to record the sensor data (see~\fig~\ref{fig:merged_frame} in Appendix~\ref{aa:perform}). We then simultaneously stream the pre-recorded sensor data (in the \ac{ul}) and send the \ac{cc} data (in the \ac{dl}) over commercial 5G network. Since part of our goal is to analyze how 5G networks affect teleoperation, to collect and analyze 5G cross-layer data, we stream the sensor data via several Samsung Galaxy S21 Ultra smartphones, tethered to the on-board computer. This is because our commercial grade 5G data collection tool, XCAL, which as of now only works with Samsung phones. 

Our methodology consists of the following steps:  \textit{Step 1): } We purchased contract SIM cards for streaming the \ac{av} data through the on-board computer via the smartphones to the AWS server. \textit{Step 2): } We then select three driving loops in Minneapolis that span 4-6~Km each and conduct several months of live \ac{av} sensor data streaming over 5G networks repeatedly over several hours spanning different time periods -- \ie morning and evening rush hours, as well as afternoons, while collecting data across all layers. \textit{Step 3): } Our experiments involve simultaneously streaming the sensor and \ac{cc} data while repeatedly driving in loops around the city center and logging at both the \ac{av} and the remote vehicle control station. Before each experiment (one complete loop in our experiments), we not only ensure proper \ac{ntp} time synchronization, but also compute the time drift between the on-board computer and the remote vehicle control server station (1000 samples) and incorporate these time differences in our analysis.

\simpletitle{Measurement Summary. }We conducted exploratory measurements over commercial 5G networks using \ac{att}, \ac{tmb}, and \ac{vz}. Altogether these measurements span about 6~months consuming 100s of GBs of 5G data and driving around the testing loops roughly about 70 times. While we studied many operators, we chose to focus on using \ac{tmb} as it was the only carrier with a primary \ac{sa} 5G deployment~\cite{carpenter2023multi,ye2023closer}.

% \vspace{-3.5ex}

\subsection{Streaming QoE Metrics}
\label{ss:qoe}

\begin{figure}[t]
  \centering
  % \vspace{-1ex}
  \includegraphics[height=15em, keepaspectratio]{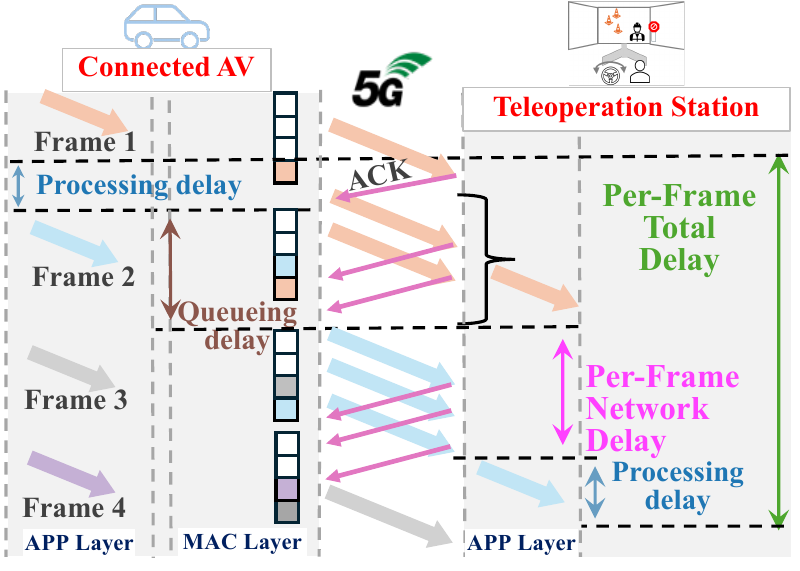}
   % \vspace{-3ex}
  \caption{Illustration of the Delay \ac{qoe} Metrics in Camera/LiDAR Streaming.}
  \label{fig:delay_illustration}
   %\vspace{-2ex}
\end{figure}

\noindent
For teleoperation, ensuring that camera (and LiDAR) data are delivered within a target latency deadline is critical in providing a human teleoperator with real-time situation awareness. Since frames are the basic units for video encoding, decoding, and playback, each frame should therefore be delivered within a target deadline (\eg 100 ms per \tbl~\ref{tab:teleop-req}). Thus, as illustrated in~\fig~\ref{fig:delay_illustration} and summarized in~\tbl~\ref{tab:qoe_metrics}, we introduce the following \emph{Per-frame} \ac{qoe} metrics to quantify the \ac{ul} (one-way) camera/LiDAR data sensor streaming performance. 

\noindent
$\bullet$ \textbf{\emph{Per-Frame (\ac{ul}) Total Delay}}: The time from when the camera/LiDAR frame is generated at the sender-side (\ie vehicle) till it is completely received and decoded at the remote teleoperation station and ready for playback.

\noindent
$\bullet$\textbf{\emph{Per-Frame (\ac{ul}) Network Delay}}: The time from when the first packet of a camera/LiDAR frame is sent from the vehicle till the last packet of this frame is received at the teleoperation station side. Note that this delay excludes the time the camera/LiDAR frame spends in the send buffer (known as ``queueing'' delay) before transmission. 

\noindent
$\bullet$ \textbf{\emph{Video Quality}}: We rely on the \ac{psnr} and \ac{ssim} to assess the video quality.

\begin{table}[t!]
\centering
\caption{Definition of \ac{qoe} Metrics Terms Used} 
 %\vspace{-1em}
\label{tab:qoe_metrics}
  \setlength{\extrarowheight}{1pt}
  \setlength{\tabcolsep}{4pt}%
  \resizebox{0.99\columnwidth}{!}{%
  \centering
  \begin{tabular}[l]{c c}
  \toprule
  \hline
  
  \begin{tabular}[c]{c@{}}
   \textbf{QoE Metrics} \\ \colorbox{spirodiscoball!80}{\textsc{\textbf{(One-Way UL)}}}
   \end{tabular}

   & \textbf{\Large Definition} 
   
  \\ \hline
    \cellcolor{gray!20} \begin{tabular}[c]{c@{}}
    \textbf{\textit{Per-Frame}} \\ \textbf{\textit{Network Delay}}

    \end{tabular}

    & \cellcolor{gray!20}
    \begin{tabular}[c]{c@{}} \textsc{\textbf{One-Way delay}} between first packet sent and \\ last packet received of a camera/LiDAR 
    frame\end{tabular}
  \\ \hline
    \cellcolor{teal!20} \begin{tabular}[c]{c@{}}
    \textbf{\textit{Per-Frame}} \\ \textbf{\textit{Total Delay}}

    \end{tabular} & \cellcolor{teal!20}
    \begin{tabular}[c]{c@{}} \textit{Per-Frame Network Delay} (\textsc{\textbf{One-Way delay}})\\ plus ($+$) ``queueing'' delay and frame \textsc{\textbf{encoding}}\\ and \textsc{\textbf{decoding}} delays\end{tabular}
  \\ \hline
   \cellcolor{purple!05} \begin{tabular}[c]{c@{}}
    \textbf{\textit{Video}} \\ \textbf{\textit{Quality}}
 
    \end{tabular} & \cellcolor{purple!05}
    \begin{tabular}[c]{c@{}} \ac{ssim} and \ac{psnr} values\end{tabular}
  \\ \hline
   \cellcolor{green!15} \begin{tabular}[c]{c@{}}
     \textbf{\textit{Perceptual Quality}} \\ \textbf{\textit{Deviation}}

    \end{tabular} & \cellcolor{green!15}
    \begin{tabular}[c]{c@{}} (\textit{SSIM} of frame received at a given playback time \\ $-$ \textit{SSIM} of the expected frame) / \textit{SSIM} of the expected frame \end{tabular}
  \\ \hline
  \bottomrule
  \end{tabular}
}
%\vspace{-4ex}

\end{table}
\noindent
$\bullet$ \textbf{\emph{Perceptual Quality Deviation}}: We introduce this metric to capture the latency-visual-quality interplay, which is defined as follows: assuming a constant latency $\delta$ determined by the time the first frame is received and displayed\footnote{If $t_1$ is the time the first frame is captured and encoded at the AV, $t_1 + \delta$ is then the time the first frame is received and displayed at the teleoperation station, and $t_i+\delta$ is the time that the $i$th frame is \emph{expected} to be displayed. If by this time, the $i$th frame is not received, the previous frame is replayed.},
 the perceptual quality deviation at the $i$th playback time is the ratio of the absolute value of the difference between the \ac{ssim} value of the \emph{actual} video frame displayed at this time and the \emph{expected} video frame (i.e., the $i$th frame) that should have been displayed to the \ac{ssim} of the $i$th frame.

For the LiDAR data, we use the term \emph{frame} to encompass the 3D point cloud data generated from a single sensor sweep of a spinning LiDAR sweep, \eg a 360$^\circ$ sweep. The per-frame total and network delay metrics also apply to LiDAR data. On the other hand, the perceptual quality deviation do not directly apply to LiDAR data. This is because, the LiDAR data is not displayed to the remote human operator. Instead, the LiDAR data will be used for 3D environment representation, object detection and recognition -- often combined with the camera data as we also analyze later in~\S\ref{ss:lidar}.

\section{Application Level Performance}
\label{s:feasibility}

Building on the streaming \ac{qoe} introduced earlier, this section examines the performance of key sensor streams over commercial 5G networks and the delay of \ac{cc} operations.

\vspace{-0.8em}

\subsection{Streaming Single Front Camera}
\label{ss:single_camera}

\noindent
We first consider streaming the front-central camera data feed -- the most crucial data feed for providing real-time situational awareness to the teleoperator. Using \ac{tmb}, we evaluate two scenarios: \textbf{Case 1: }We stream the raw MJPEG camera feed using \ac{rtsp}. \textbf{Case 2: }To evaluate the effectiveness of data compression for teleoperation, we consider several compression schemes and stream the compressed camera feed using both \ac{rtsp} and \ac{webrtc}. We repeat these experiments multiple times in our loops.

\begin{figure*}[htbp]
    \centering
    
    \begin{subfigure}[b]{0.2\textwidth} 
        \centering
        \includegraphics[width=\textwidth,  keepaspectratio]{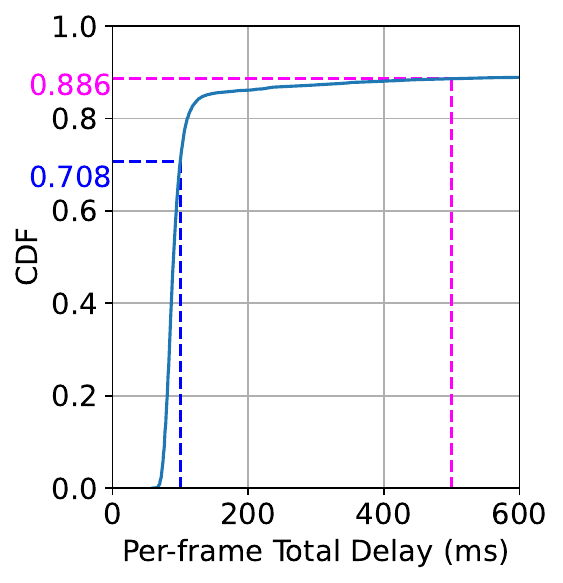}

        \caption{[\ac{ul}] \emph{Per-Frame Total Delay}}
        \label{fig:single_per_frame_delay}
    \end{subfigure}
    \hfill
    \begin{subfigure}[b]{0.28\textwidth} 
        \centering
        \includegraphics[width=\textwidth, keepaspectratio]{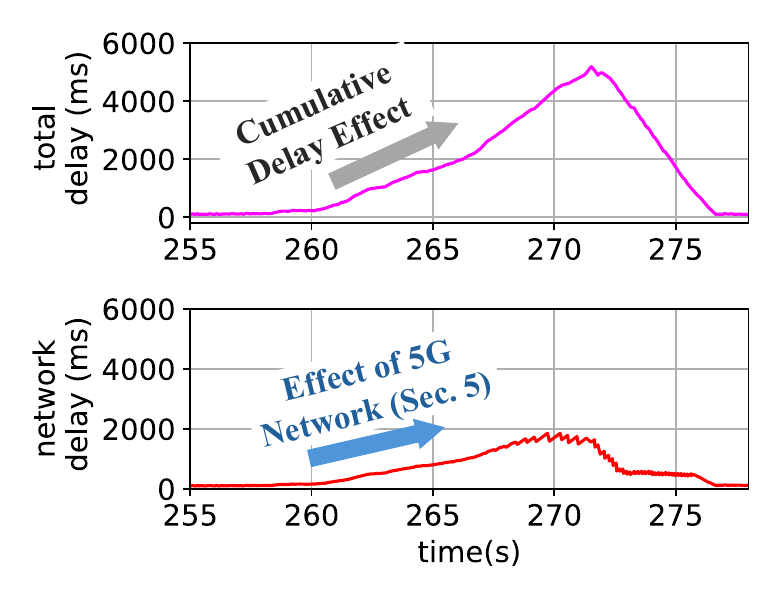}

        \caption{[\ac{ul}] \emph{Per-Frame Total} and \emph{Per-Frame Network Delay}}
        \label{fig:time-series-delay-plot-single}
    \end{subfigure}
    \hfill
    \begin{subfigure}[b]{0.19\textwidth} 
        \centering
        \includegraphics[width=\textwidth, keepaspectratio]{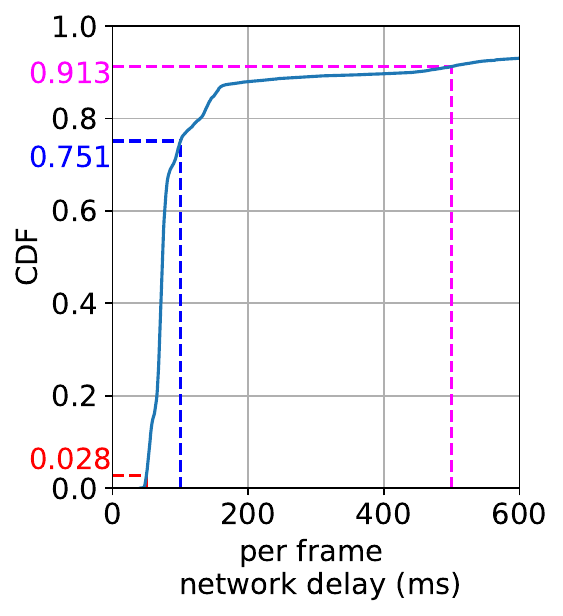}

        \caption{[\ac{ul}] \emph{Per-Frame Network Delay}}
        \label{fig:network-delay-mjpeg-single}
    \end{subfigure}
    \hfill
    \begin{subfigure}[b]{0.29\textwidth} 
        \centering
        \includegraphics[width=\textwidth,keepaspectratio]{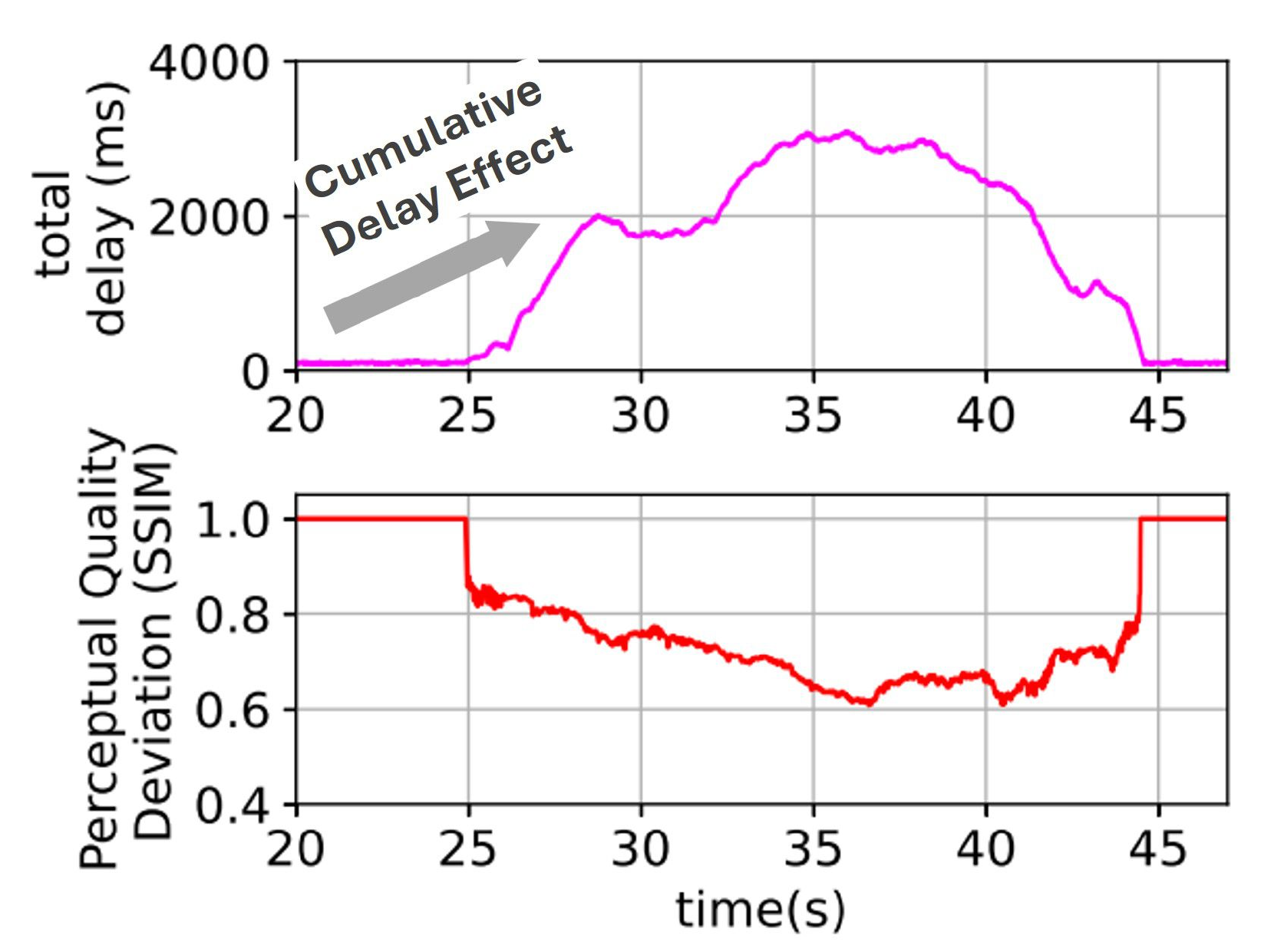}
        \caption{[\ac{ul}] \emph{Per-Frame Total Delay} and \emph{Perceptual Quality Deviation}}
        \label{fig:percetual-deviation-mjpeg-single}
    \end{subfigure}
    \vspace{-2ex}
    \caption{\ac{qoe} Metrics When Streaming the Raw Front-Central Camera -- the Most Crucial Data Feed Over \ac{tmb} 5G}
    \label{fig:qoe_comparison_single_video}

\end{figure*}

\begin{figure*}[htbp]
    \begin{minipage}[c]{0.6\textwidth}
            \begin{subfigure}[c]{0.33\textwidth} 
                \centering
                \includegraphics[width=\textwidth, height=1.4in]{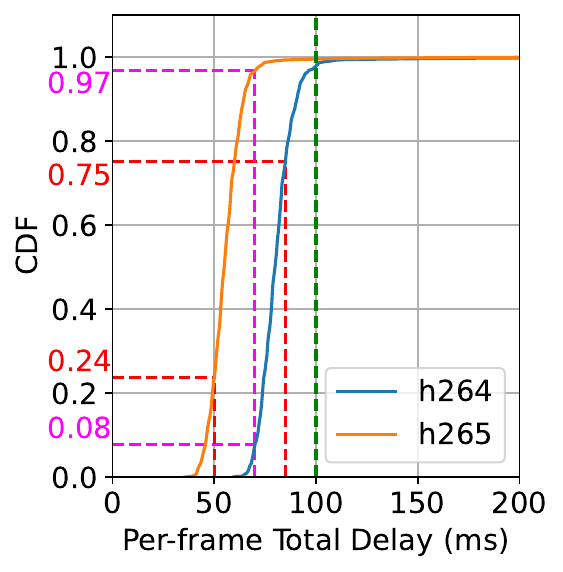}
                \caption{[\ac{ul}] \emph{Per-Frame Total Delay} for I Frames}
                \label{fig:total-delay-h264-h265}
            \end{subfigure}
            \hfill
            \begin{subfigure}[c]{0.33\textwidth}
                \centering
                \includegraphics[width=0.97\textwidth, height=1.4in]{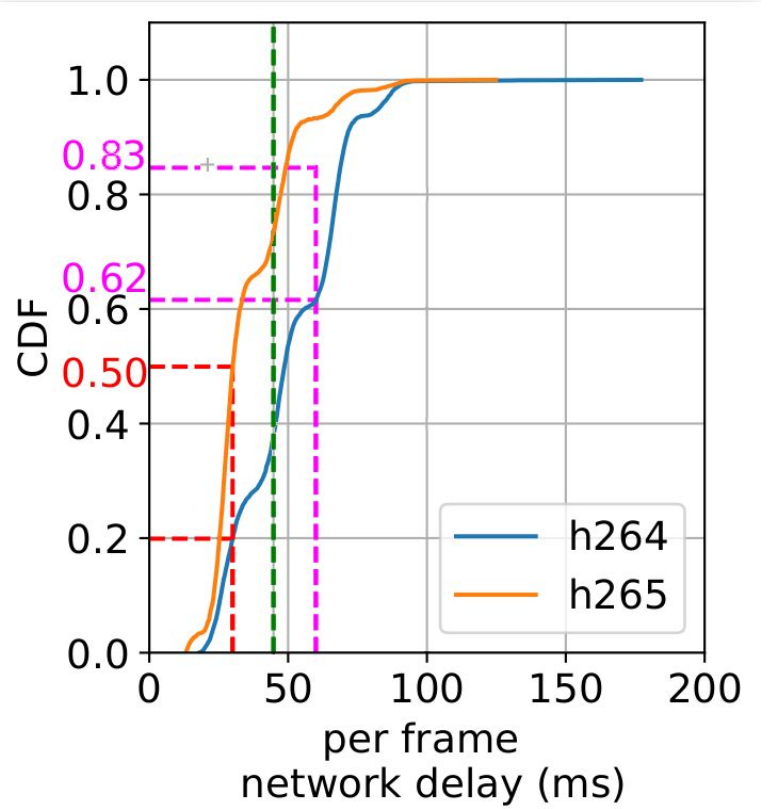}
                \caption{[\ac{ul}] \emph{Per-Frame Network Delay} for I Frames}
                \label{fig:network-delay-h264-h265}
            \end{subfigure}
            \hfill
            \begin{subfigure}[c]{0.32\textwidth}
                \includegraphics[width=\textwidth, height=1.4in]{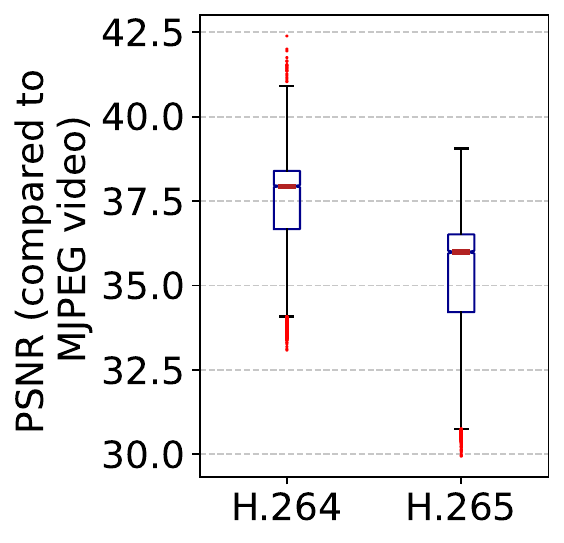}
                \caption{\emph{Video Quality} Indicated with \ac{psnr}}
                \label{fig:video-quality-h264-h265}
            \end{subfigure}
            \vspace{-2ex}
        \caption{Front-Camera Streaming H.264 and H.265 Performance with \ac{rtsp}}
        \label{fig:rtsp_single}
    \end{minipage}
    \hfill
    \begin{minipage}[c]{0.38\textwidth} 
            \begin{subfigure}[b]{0.48\textwidth}
                \centering
                \includegraphics[width=0.97\textwidth, height=1.4in]{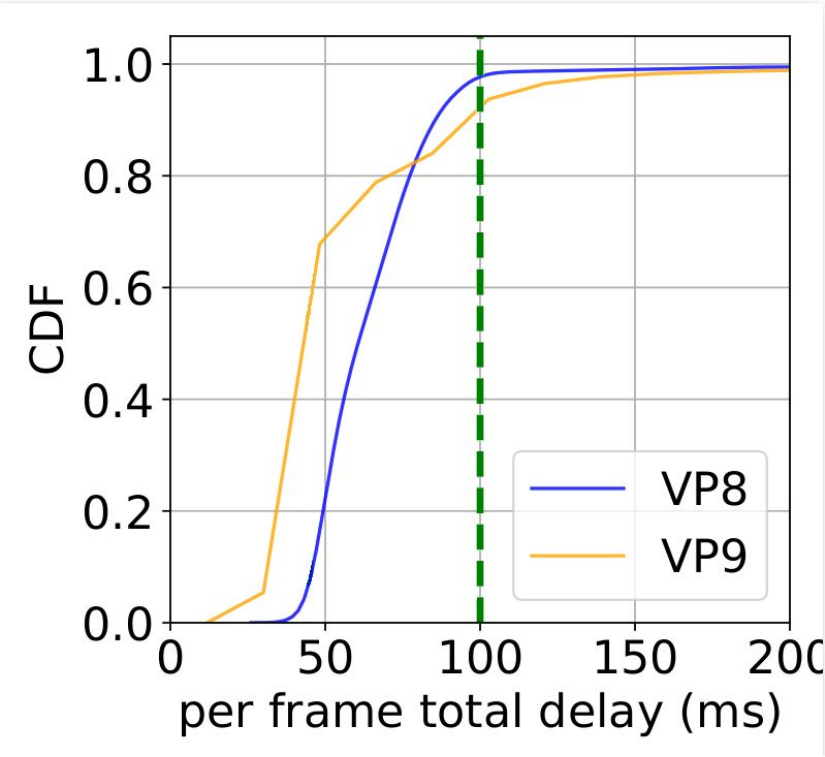}
                \caption{[\ac{ul}] \emph{Per-Frame Total Delay}}
                \label{fig:webrtc_single_totaldelay}
            \end{subfigure}
                 \hfill
            \begin{subfigure}[b]{0.48\textwidth}
                \centering
                \includegraphics[width=0.98\textwidth, height=1.4in]{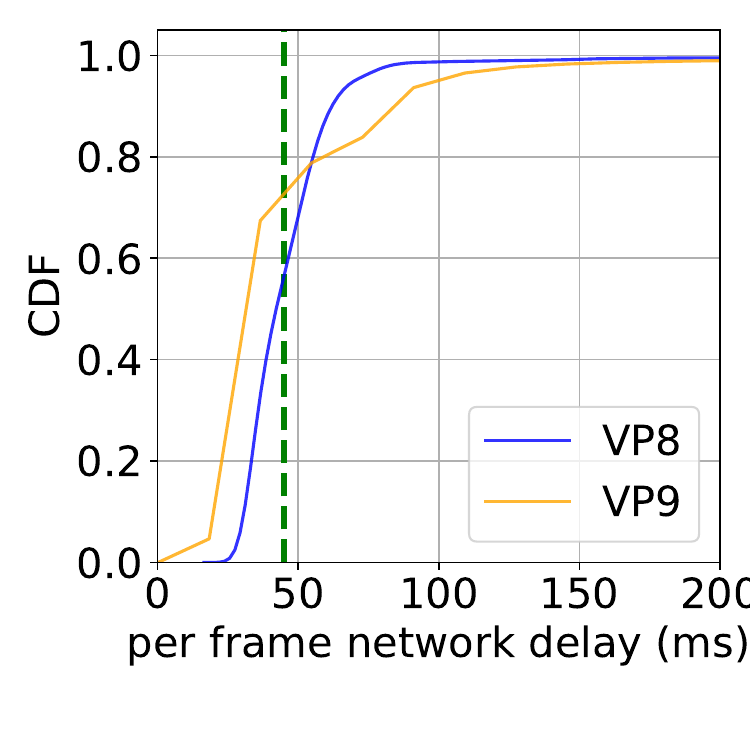}
                \caption{[\ac{ul}] \emph{Per-Frame Network Delay}}
                \label{fig:webrtc_single_networkdelay}
            \end{subfigure}
            \vspace{-2ex}
        \caption{Front-Camera Streaming VP8 and VP9 Performance with \ac{webrtc}}
        \label{fig:webrtc_single}
    \end{minipage}
    \vspace{-2ex}
\end{figure*}

\simpletitle{Case 1: Streaming Raw Camera Feed. } Recall from ~\S\ref{s:capabilities} that, for teleoperation to be feasible, the \ac{etoe} \ac{ul} delay, \ie the \emph{Per-Frame Total Delay} should be below 100~ms. We consider the base case (\ie streaming the front-central camera feed) and quantify the \emph{Per-Frame Total Delay} to understand the feasibility of teleoperation over commercial 5G today. 

\textbf{\emph{(\ac{ul}) Per-Frame Total Delay}.} \fig~\ref{fig:single_per_frame_delay} shows the \emph{(\ac{ul}) Per-Frame Total Delay}. We make the following observations: (1)~The minimum \emph{Per-Frame Total Delay} exceeds 50~ms, with 29.2\% of frames surpassing the 100~ms deadline -- of which 11.4\% experience delays greater than 500~ms. (2)~Notably, the \emph{Per-Frame Total Delay} progressively worsens over time (\ie  accumulates) -- a phenomenon we refer to as the \emph{cumulative delay effect} (see~\fig~\ref{fig:time-series-delay-plot-single} top figure). This is because the \ac{av}'s camera data begins to queue in the \ac{ul} buffer. (3)~The queuing is triggered by a domino effect due to the 5G network's inability to transmit data to the teleoperation station. This effect is captured by a corresponding increase in \emph{Per-Frame Network Delay}, also illustrated in~\fig~\ref{fig:time-series-delay-plot-single} bottom plot. The \emph{Per-Frame Network Delay} (partially) provides a clearer justification for the critical role of 5G networks in teleoperation.

\textbf{\emph{(\ac{ul}) Per-Frame Network Delay}. } ~\fig~\ref{fig:network-delay-mjpeg-single} shows the median \emph{Per-Frame Network Delay} is around 73.5~ms, which is 28~ms higher than the maximum 5G network delay threshold required for teleoperation (see~\tbl~\ref{tab:teleop-req}). However, the distribution has a long tail, with 72.3\% of the frames with delay between 50 and 100~ms, 16.2\% with delay between 100~ms and 500~ms, and 8.7\% with a network delay larger than 500~ms. 

\textbf{\emph{Perceptual Quality Deviation}. } Lastly, we quantitatively understand the \emph{Perceptual quality deviation} in~\fig~\ref{fig:percetual-deviation-mjpeg-single} -- The \emph{cumulative delay effect} has a proportionally negative impact on the \emph{Perceptual quality deviation}. In other words, the video frames are progressively delayed, degrading the operator’s ability to maintain situational awareness and potentially causing confusion, which can lead to delayed reactions from the teleoperator.

\simpletitle{Case 2: Streaming Compressed Camera Feed. }Now we ask the question, \emph{can video data compression techniques help teleoperation?} -- In other words, \emph{can compression help favor the majority of the frames to arrive within the latency requirements for Teleoperation?} \emph{What are the tradeoffs?} To answer these questions, we configure H.264 and H.265 with \ac{rtsp} and VP8\cite{rfc6386-vp8} and VP9\cite{ietf-payload-vp9-16} in \ac{webrtc} to stream the front-camera data. \fig~\ref{fig:rtsp_single} and \fig~\ref{fig:webrtc_single} show the latency performance with \ac{rtsp} and \ac{webrtc} respectively. Generally, compression reduces both the \emph{Per-Frame Total} and \emph{Per-Frame Network} Delays. For instance, with H.265 and \ac{rtsp}, only 0.487\% of the frames experience a \emph{Per-Frame Total Delay} greater than 100~ms, and 17\% of the frames experience a \emph{Per-Frame Network Delay} exceeding 45~ms, the target latency requirements for teleoperation. Nonetheless, such tail latency performance is still not ideal, as a critical safety situation that requires timely human teleoperator intervention may occur during a concentrated period of the tail events with bad 5G network conditions. Additionally, we see that increasing compression reduces delay at the expense of lower video quality (see~\fig~\ref{fig:video-quality-h264-h265}). Notably, the median \emph{Per-Frame Total} and \emph{Per-Frame Network} Delays achieved with \ac{webrtc} are 30.47\% and 50.81\% lower when compared with \ac{rtsp}. This is due to the effectiveness of the network adaptation and congestion control mechanisms present in \ac{webrtc} but absent in \ac{rtsp}, which is the main focus of our investigations later in~\S\ref{s:cc-adaptation}.

\subsection{Streaming Merged Cameras}
\label{ss:merge_camera}

\noindent
To provide complete real-time situational awareness, it is not sufficient to only stream the data from the front-view camera; the side-views provided by the left-front and right-front cameras are also important, especially when the vehicle needs to change lanes or make turns. We therefore also consider the streaming performance of the merged video streams from all three cameras (see~\fig~\ref{fig:merged_frame} in Appendix~\ref{aa:perform}). Intuitively, the \emph{Per-Frame Total} and \emph{Per-Frame Network} Delays when streaming data from 3 cameras will be worse than the single camera, even with compression. For completeness, we include these results for both \ac{rtsp} and \ac{webrtc} in~\fig~\ref{fig:merged_cameras_streams} and \fig~\ref{fig:merged_video_webrtc}  in Appendix~\ref{aa:perform}. Overall, for \ac{rtsp}, 45\% of the frames violate the 45~ms \emph{Per-frame Network} delay deadline and 13\% violate the 100~ms \emph{Per-Frame Total} delay deadline. While for \ac{webrtc}, 48.31\% of the frames will violate the \emph{Per-Frame Network} Delay and 3.64\% violate the \emph{Per-Frame Total} delay requirement.

\subsection{LiDAR Data Streaming}
\label{ss:lidar}

\noindent
To stream the LiDAR data, we built a custom client-server UDP-based application that packetizes the \ac{ros} bag data for streaming over 5G. In these experiments, we use the term \emph{frame} to encompass the 3D point cloud data generated from a single round of LiDAR sweep, \eg a 360$^\circ$ sweep, and quantitatively analyze the \emph{Per-Frame Network Delay}. As shown earlier in~\tbl~\ref{tab:av-req}, the \ac{ul} throughput required to stream the 64 and 128 beams LiDAR data is 277~Mbps and 307~Mbps, respectively. Given these throughput requirements, we further explore two approaches to reduce LiDAR data before streaming over 5G. 

\noindent
\simpletitle{1) Voxel-based Downsampling. }Using downsampling with voxel sizes of $0.1^3$ and $0.5^3$ (cubic meters) of the 64 beams LiDAR data, we can reduce the throughput requirements from  277~Mbps down to about 121.4~Mbps and 50.9~Mbps, respectively -- the former is still significantly higher than the average \ac{ul} throughput of the three operators. In contrast, the latter is lower than that of \ac{tmb} and \ac{vz}, but still slightly above that of \ac{att}. 

\noindent
\simpletitle{2) LiDAR Compression. }We chose Google's Draco~\cite{githubGoogledraco} because it can reduce the LiDAR data to a third of its original size, and is faster than other LiDAR compression tools we have tested, like Octree~\cite{Octree}.

\begin{figure}[t]
\begin{minipage}{0.23\textwidth}%
\centering
    \includegraphics[scale=0.46, keepaspectratio]{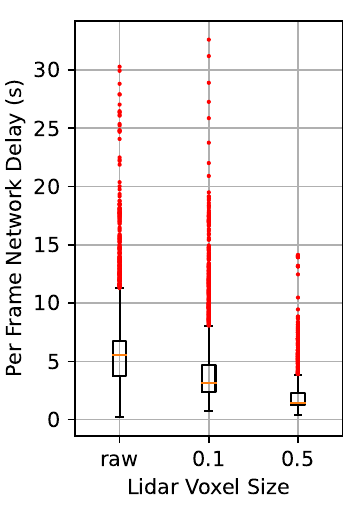}
   \vspace{-2ex}
  \caption{[\ac{ul}] \emph{Per-Frame Network Delay} of LiDAR.}
  \label{fig:network_delay_single_lidar}
    
\end{minipage}
\hfill
\begin{minipage}{0.23\textwidth}
    \centering
    \includegraphics[scale=0.39, keepaspectratio]{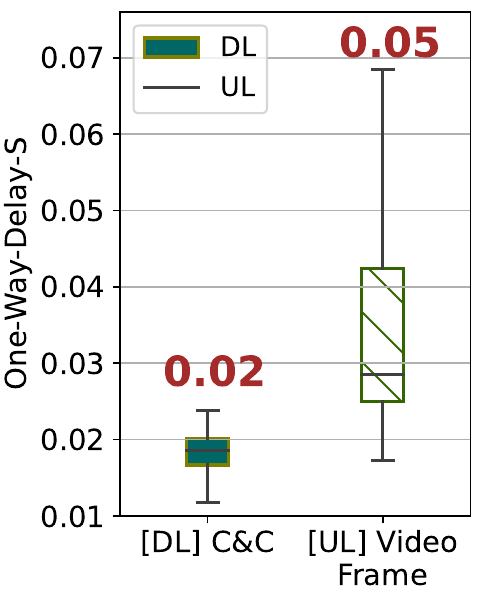}
    \vspace{-2ex}
    \caption{[\ac{dl}] \ac{cc} vs. [\ac{ul}] Video Frame Delay.}
    \label{fig:cnc_frames}
    
\end{minipage}
 \vspace{-3ex}
\end{figure}

\begin{figure*}[htbp]
  \centering
  \includegraphics[width=0.9\textwidth, keepaspectratio]{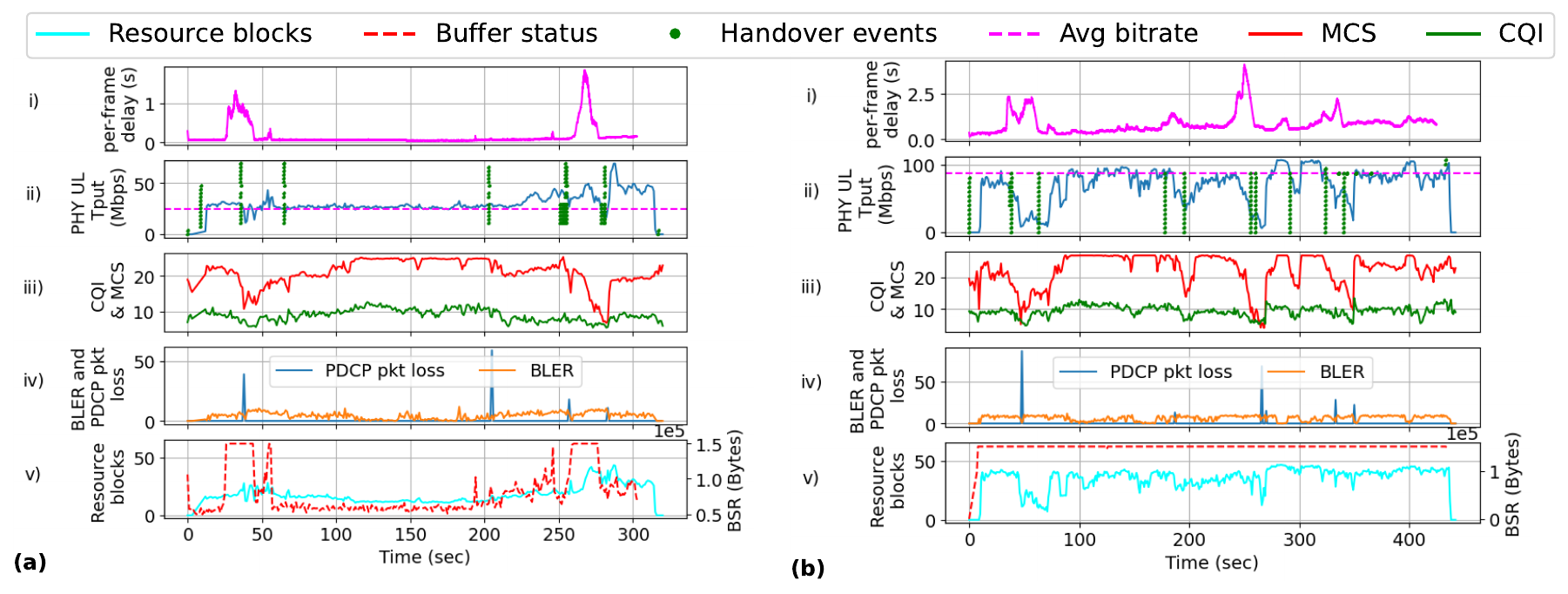}
   \vspace{-3ex}
  \caption{5G Impact on \emph{Per-Frame (\ac{ul}) Network Delay} for Teleoperation. Using  \ac{rtsp} to Stream:\\ (a) Single (front) Camera, and (b) Merged (front, left, \& right) Cameras}

  \label{fig:trace:single-video}
 %\vspace{-2ex}
\end{figure*}

\noindent
\simpletitle{Results. }We find the \emph{Per-Frame Network Delay} of the 64 beams LiDAR data streaming is very poor, even with downsampling, as shown in~\fig~\ref{fig:network_delay_single_lidar}. With $0.5^3$ voxel size, the median network delay is 2~seconds, whereas for one raw frame, the median network delay is about 6~seconds, making it nearly infeasible to stream high-resolution LiDAR data in real-time over 5G networks. We omit the results for streaming 128 beams and multiple LiDAR streams, as they will obviously perform worse. With Draco, although we can significantly reduce the LiDAR data to one-third of its original size, the caveat is that the average compression and decompression times are approximately 47~ms and 15~ms, respectively. In contrast, voxel downsampling takes about 14~ms with no processing required at the remote teleoperation station. The large additional processing overhead associated with Draco, combined with network delays, makes it less suitable for AV teleoperation. Further downsampling of LiDAR data is not beneficial, as it leads to performance degradation in downstream AI tasks, such as object detection and recognition. For interested readers, we provide a detailed discussion of the impact of compression for downstream AI tasks for \ac{av} in Appendix~\ref{aa:ai_tasks}.

\vspace{-0.5em}

\subsection{Command \& Control (C\&C)}
\label{ss:cc}
As shown in \fig\ref{fig:setup}, the test vehicle used in this study is equipped with a DBW (Drive-By-Wire) system, which enables it to read and write CAN (Controller Area Network) messages. The operator remotely controls the vehicle using the Logitech simulator platform at the remote station. Communication between the simulator and the vehicle is established through \ac{grpc} communication protocol, ensuring efficient and low-latency data transmission over the 5G network. gRPC's communication model achieves low-latency data transmission by minimizing the use of extensive error-checking mechanisms and connection handshakes. This makes it particularly well-suited for vehicle \ac{cc} transmissions, where real-time responsiveness is prioritized over guaranteed delivery. Control commands, including steering, acceleration, and braking, transmitted from the remote station were first converted into low-level commands on the vehicle side before being executed by the vehicle’s control systems (See Appendix~\ref{aa:cnc} for more details). \fig~\ref{fig:cnc_frames} shows the \ac{dl} \ac{cc} compared to the \ac{ul} \emph{Per-frame Network} delays. As per the teleoperation delay requirements in \tbl~\ref{tab:teleop-req}, we find that 64.29\% and 37.20\% of the \ac{cc} messages were delivered within the application and network requirements, respectively, with a median delay of {17.29}~ms.\\

\noindent

\simpletitle{\textbf{\textit{Implications \& Key Takeaways.}}} Our analysis reveals that streaming a single camera's data over a commercial 5G network is feasible. However, situational awareness requires streaming multiple cameras and LiDAR data, which renders teleoperation infeasible. While compression improves feasibility, it degrades visual quality and impairs downstream AI tasks (See~Appendix~\ref{aa:ai_tasks}). Additionally, analysis of the streaming \ac{qoe}s highlights the critical role of 5G in teleoperation. We next examine 5G’s impact on ToD in details.

\section{5G Impact on AV Teleoperation}
\label{s:5gRAN}

Minimizing delay is crucial for teleoperations. However, as shown earlier, \emph{Per-frame network delay} often exceeds the 45~ms or 100~ms requirements (see~\tbl~\ref{tab:teleop-req}) and is unstable. We analyze the 5G dynamics to identify the causes.

\subsection{What 5G Factors Affect Teleoperation?}
\label{ss:5g_phy}

As the \ac{av} moves around, the \ac{bs} (\ie \ac{gnb}) needs to allocate network resources for \ac{ul} sensor data transmissions. The \ac{gnb} considers several factors when allocating resources. Among many others, channel condition assessments play a crucial role. The \ac{av} reports its channel conditions to the network using a Channel Quality Indicator (CQI) value, which ranges from 1 to 15, where 15 indicates excellent channel conditions. The network, in turn, uses this \ac{cqi} value to determine which \ac{mcs} to use for the impending transmission [See $\S$5.1.6 in~\cite{3GPP_TS38.214v16.2}]. Generally, a high \ac{cqi} value results in a higher \ac{mcs} value -- \emph{if} there is sufficient data buffered to warrant it~\cite{rossPam}. This effect is evidenced by the \ac{bsr} of the \ac{av} being always full. Additionally, \acp{hos} are triggered to switch between \acp{BS} when the \ac{av} moves around, which can result in packet loss and retransmissions -- \ie \acp{blers}. 

\noindent
\simpletitle{5G \emph{Cross-layer} and \emph{\ac{etoe}} Correlation Analysis. }To visually illustrate the impact of 5G on \emph{Per-frame network delay}, in \fig~\ref{fig:trace:single-video}(a) and \fig~\ref{fig:trace:single-video}(b), we show the time-series plots of the 5G dynamics, cross-correlating them with the \ac{av} \ac{qoe} \emph{Per-frame network delay} for about 7.5 minutes while driving in a loop and streaming a \emph{single front} camera and \emph{merged} camera feeds respectively over \ac{tmb}'s \ac{sa} network. The first row i) and the second row ii) plot the time series of \emph{Per-frame network delay} and 5G \ac{ul} \ac{phy} throughput, respectively, with \ac{ho} events represented using green dotted lines in the second row. Rows iii) to v) show several key 5G parameters: iii) \ac{ul} \ac{cqi} and \ac{mcs};  iv) PHY \acp{blers} and \ac{pdcp} loss rates; and v) the \ac{ul} \acp{rb} allocated and the \ac{av} \ac{bsr} (right y-axis). 

\begin{figure*}[t]
    \begin{minipage}[c]{0.29\textwidth}%
        \centering

        \includegraphics[width=0.94\linewidth, keepaspectratio]{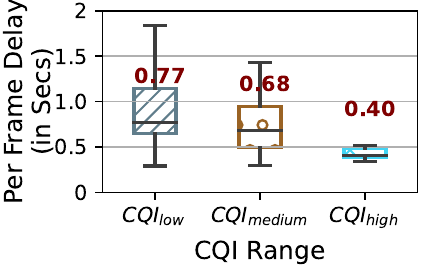}
        \vspace{-1.0em}%
        \caption{CQI Impact on \emph{Per-frame Network Delay}}
        \label{fig:CQI_impact}
    \end{minipage}
    \hfill%
    \begin{minipage}[c]{0.29\textwidth}%
        \centering
        \includegraphics[width=0.97\linewidth, keepaspectratio]{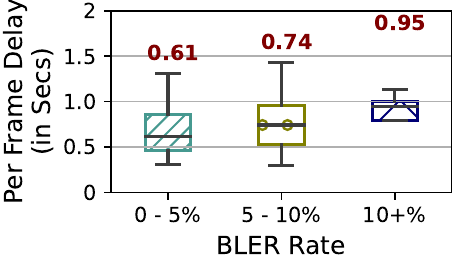}
        \vspace{-1.0em}%
        \caption{BLER Impact on \emph{Per-frame Network Delay}}
        \label{fig:bler_impact}
    \end{minipage}%
    \hfill
    \begin{minipage}[c]{0.38\textwidth}%
        \centering
        \includegraphics[width=0.95\linewidth, keepaspectratio]{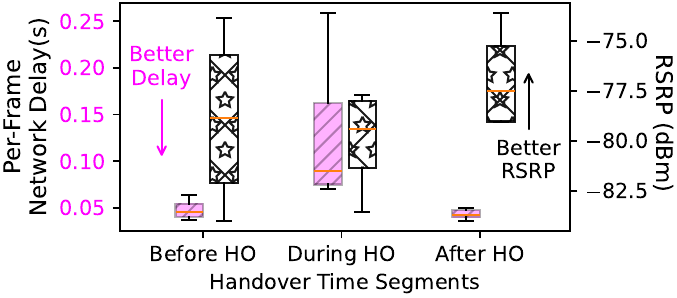}
        \vspace{-1em}%
        \caption{\acp{hos} Impact on \emph{Per-frame Network Delay}}
        \label{fig:hos_impact}
    \end{minipage}

   % \vspace{-3ex}% 
\end{figure*}

We make the following observations: (1)~When the PHY layer throughput drops dramatically (e.g., 250 -280 secs in plot ii) of \fig~\ref{fig:trace:single-video}(b)), the \emph{Per-frame network delay} increases significantly (from 0.1~seconds to 4~seconds), and can last more than 30~seconds. (2)~\acp{hos} typically occur in regions where there is a noticeable spike (\ie increase) in the \emph{Per-frame network delay}. Not all \acp{hos} cause significant disruption; some cause extended periods of high delay, with values ranging from hundreds of milliseconds to several seconds, and occasionally lasting tens of seconds. A particularly interesting observation is the compounded impact of multiple \acp{hos} occurring within a short time window. For instance, in plot ii) of \fig~\ref{fig:trace:single-video}(a) around 250 seconds and in plot ii) of \fig~\ref{fig:trace:single-video}(b) around 270 seconds, the compounded effect of several \acp{hos} results in much more severe delays than the effect of a single \ac{ho} during the same period -- this is a key point that we explore further in~\S\ref{ss:retxs_hos}. (3)~\acp{blers} also increase \emph{Per-frame network delay}, although the effects are less pronounced, as we quantify this later (see~\S\ref{ss:retxs_hos}). \acp{hos}, and to a lesser extent, \acp{blers}, can lead to the \ac{av} \ac{mac} buffer build-up, as indicated by the buffer status (red curve, right y-axis) in plot v).  (4)~In single camera streaming, except in the beginning and the end, the number of \acp{rb} reaches 20 per second. When the \ac{av} \ac{mac} buffer builds up, more \acp{rb} are allocated as expected -- despite poor channel conditions (\ie low \ac{cqi}/\ac{mcs}). In contrast, the merged cameras require significantly higher bit rate, the \ac{av} \ac{mac} buffer consistently remains full, necessitating the allocation of at least 32 \acp{rb} (see plot v) in~\fig~\ref{fig:trace:single-video}(b)).

The above \emph{cross-layer} correlation analysis establishes a clear connection between the 5G network dynamics impact on the \ac{qoe} of \tod. However, the underlying causes of \tod performance, and the ability to quantitatively establish a causal relationship between key 5G factors and the \ac{av} \ac{qoe} is yet to be fully explored. Next, we aim to assess and quantify the precise impact of these factors.

%\vspace{-0.5em}

\subsection{Quantitative Analysis of the Impact of Key 5G Factors on AV Teleoperation}
\label{ss:retxs_hos}

\simpletitle{Radio Resource Allocations.} The radio resource allocation does not have a \emph{direct} effect on the observed \emph{Per-frame network delay} (\fig~\ref{fig:trace:single-video}(a) \& \fig~\ref{fig:trace:single-video}(b)). Particularly, channel conditions (\ie \ac{cqi}) do not appear to affect the number of \ac{ul} \acp{rb} allocated by the \ac{bs}, in contrast to the \ac{mac} buffer status. \acp{hos}, on the other hand, do sometimes reduce the number of \acp{rb} allocated -- this is not too surprising. However, when multiple \ac{av}s simultaneously stream video feeds, competition for radio resources affects the number of \acp{rb} allocated per vehicle, therefore affecting the \emph{Per-frame network delay}, as we later analyze in~\S\ref{s:e2eApproaches}.

\simpletitle{5G Channel Conditions. }To quantitatively understand how \ac{cqi} impacts \emph{Per-frame (\ac{ul}) network delay}, we stream the merged video camera data, which fills the \ac{ul} buffer (to minimize the impact of data on the network delay) and quantify the direct effect of \ac{cqi}. We categorize the \ac{cqi} values into three bins: \cqil = (6, 9] (``poor''), \cqim = (9, 12] (``fair''), and \cqih = (12, 15] (``good''). \fig~\ref{fig:CQI_impact} shows the box plot of the \emph{Per-frame network delay} for each \ac{cqi} bin. We observe that under poor channel conditions, the average \emph{Per-frame network delay} is 770~ms, which is significantly higher than the 400~ms observed under good conditions, representing a 92.5\% increase.

\simpletitle{BLERs. }The 5G PHY/\ac{mac} employs the \ac{harq} to  recover from bit errors and failed transmissions. \fig~\ref{fig:bler_impact} shows the effect of \ac{bler} on the \emph{Per-frame network delay} for a single camera streaming. The delay increases by about 55.7\% when transitioning from 0-5\% \ac{bler} to 10+\% \ac{bler}, rising from 0.61~seconds to 0.95~seconds. This increase is due to the higher number of PHY/\ac{mac} retransmissions.

\simpletitle{Handovers (\acp{hos}). }While previous works \cite{hassan2022vivisecting, ghoshal-imc23, xu2020understanding} have shown that \acp{hos} negatively affect the throughput and latency, our goal here is to quantitatively understand their impact on \ac{av} teleoperations, specifically the \ac{ul} \emph{Per-frame network delay}. In the case of intra-RAT (radio access technology) \acp{hos} (\ie 5G $\rightarrow$ 5G), we focus on A3 \acp{hos} observed 97.03\% of the times in our experiments; (1) A3 \acp{hos} -- defined as the \ac{rsrp} of the neighboring cell becomes an offset better than the serving cell. (2) A3 Ping-Pong \acp{hos} -- defined as, multiple \acp{hos} between cell ids $PCI^{n-1}$, $PCI^{n}$, and $PCI^{n+1}$ happening within a short time window, 15~seconds in our analysis in which $PCI^{n-1} = PCI^{n+1}$.

\textbf{A3 \acp{hos} Impact on \ac{av}. }For A3 \acp{hos}, we are interested in understanding the \emph{Per-frame network delay} (5~seconds) before, during, and (5 seconds) after a \ac{ho}. Notice that the $\Delta$\emph{Per-frame network delay} before and after a \ac{ho} quantifies the improvement due to a \ac{ho}. \fig~\ref{fig:hos_impact} shows the \emph{Per-frame network delay} (left axis) and the corresponding \ac{rsrp} values before, during, and after a \ac{ho} without \acp{hos} failures. See \S\ref{s:cc-adaptation} later for the effect of \ac{ho} failures. Not surprising, $\Delta$\ac{rsrp} defined as $RSRP_{after} - RSRP_{before}$ a \ac{ho} is positive. This is a direct consequence of the \ac{ho} decision. Importantly, although the result of \acp{hos}  (\ie after) generally improve the \emph{Per-frame network delay} by about 7.83\% in our experiments (going from 0.046~secs to 0.0424~secs), their impact (\ie during) is far more detrimental, causing the \emph{Per-frame network delay} to increase (\ie worsen) by about 86.04\% (going from 0.048~secs up to 0.0893~secs). 

\begin{figure}[t!]
    \vspace{-3.5ex}
        \centering
        \includegraphics[scale=0.44, keepaspectratio]{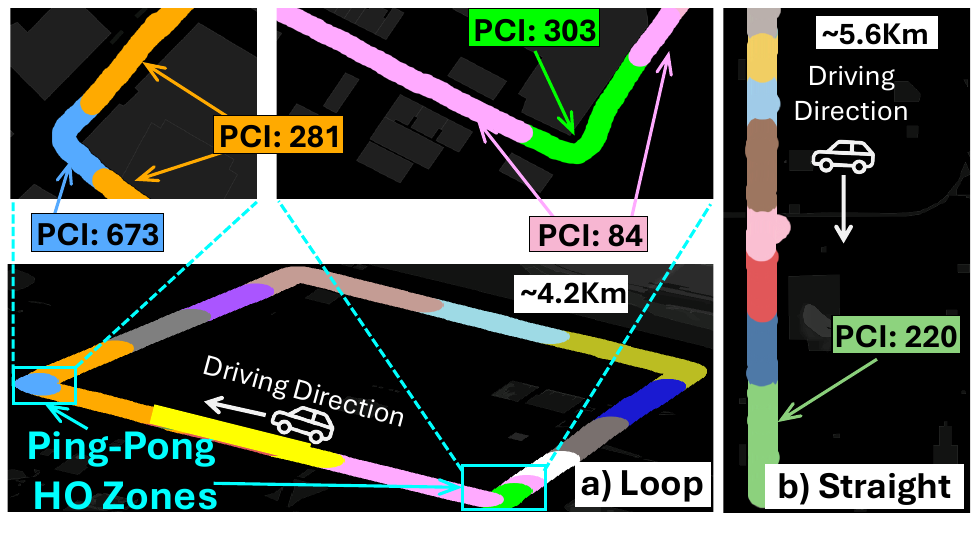}
        \vspace{-1ex}% 
        \caption{Ping-Pong \acp{hos} occurring while driving in a loop and making turns.}
        \vspace{-2.5ex}% 
        \label{fig:pp_hos}

\end{figure}

\textbf{Ping-pong \acp{hos} Impact on \ac{av}. }We find that ping-pong \acp{hos} generally occur when driving in a loop, particularly when turning. To illustrate, we compare \acp{hos} when driving in a loop and on a straight line. \fig~\ref{fig:pp_hos} shows the cell IDs (\ie PCIs) with unique colors the \ac{av} is connected to when driving in a loop ($\approx$4.2Km) compared to a straight line ($\approx$5.6Km). Notice that, unlike when driving on a straight line, there are zones of ping-pong \acp{hos} (\ie unnecessary \acp{hos}) when the \ac{av} is turning. For instance, the top-zoomed plot on the left of \fig~\ref{fig:pp_hos}a) shows that the PCIs the \ac{av} is connected to changes from PCI:281 $\rightarrow$ PCI:673 $\rightarrow$  PCI:281 within a short time window (15 seconds). The compounded effect of this pong-pong \acp{hos} causes a 56-85\% in the \emph{Per-frame network delay}, presenting a significant problem for \ac{av} teleoperations.

\begin{figure}[t!]
    \vspace{-1ex}%
        \centering
        \includegraphics[scale=0.44, keepaspectratio]{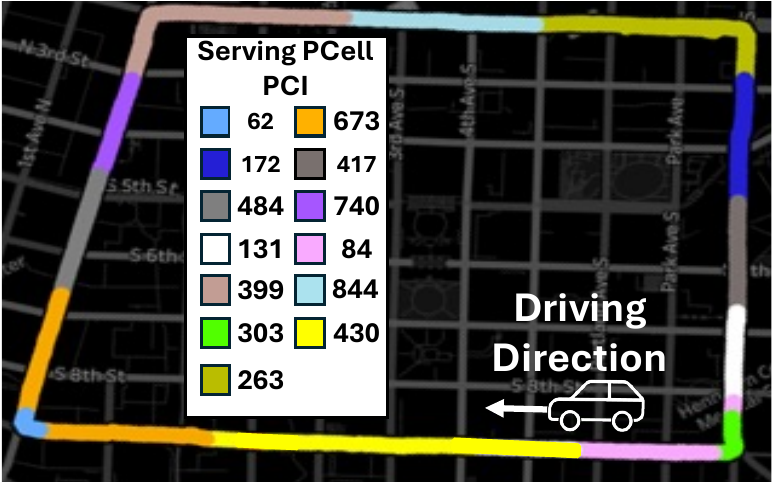}
        % \vspace{-2.0em}%
       \vspace{-2.5ex}% 
        \caption{Full Drive Loop, showing the density of PCIs.}
        % \vspace{-3ex}% 
        \label{fig:drive_loop_pcis}
\end{figure}

\noindent
\simpletitle{\textbf{\textit{Implications \& Key Takeaways. }}}  Our analysis points to key 5G factors—\acp{hos} (unnecessary ping-pong \acp{hos} \& A3 \acp{hos}), 5G channel conditions, and \ac{bler}—which need to be addressed for \ac{av} teleoperation on commercial 5G networks. 

Practically, one could leverage vertical application information like the vehicle location, speed, and trajectory to i) design a \emph{proactive, trajectory-driven} \acp{hos} mechanism to minimize ping-pong \acp{hos}, particularly when turning, and ii) implement a \emph{mobility-aware} link adaptation (\eg \ac{mcs} selection) mechanism to reduce \ac{bler}. Such approaches could make \ac{av} teleoperation feasible. Also, understanding adaptation strategies at both the network and application layers could be essential -- the major focus of our next section. 

\begin{figure*}[t!]
    \begin{minipage}[c]{0.57\textwidth}%
        \centering
        \includegraphics[scale=0.44, keepaspectratio]{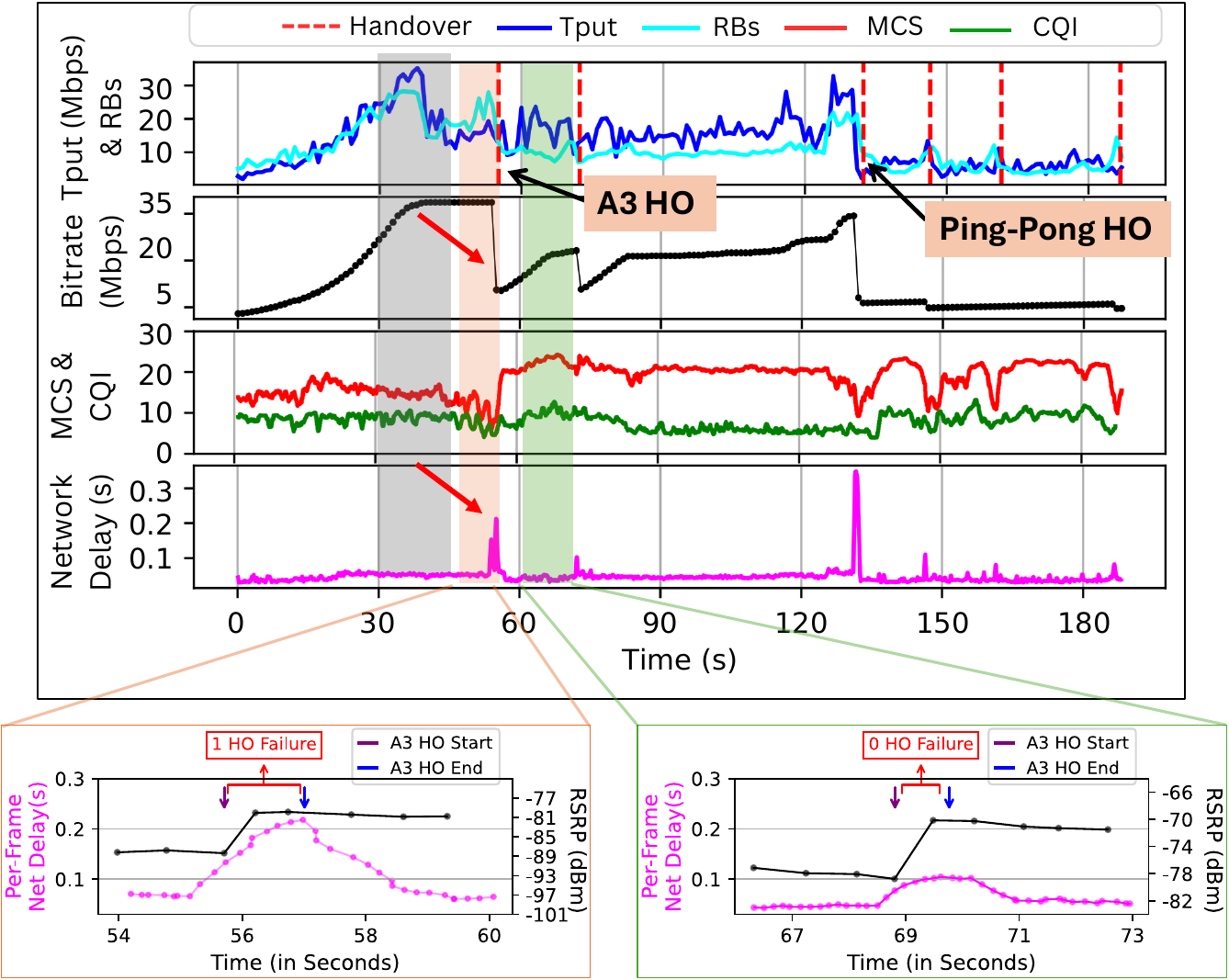}
        %\vspace{-2ex}
        \caption{Cross-Layer Analysis of WebRTC Single Camera Streaming: 5G \ac{phy} Impact on App. Layer Behavior.}
        \label{fig:multi-layer-webrtc}
    \end{minipage}
    \hfill
    \begin{minipage}[c]{0.42\textwidth}
        \centering
        \includegraphics[scale=0.55, keepaspectratio]{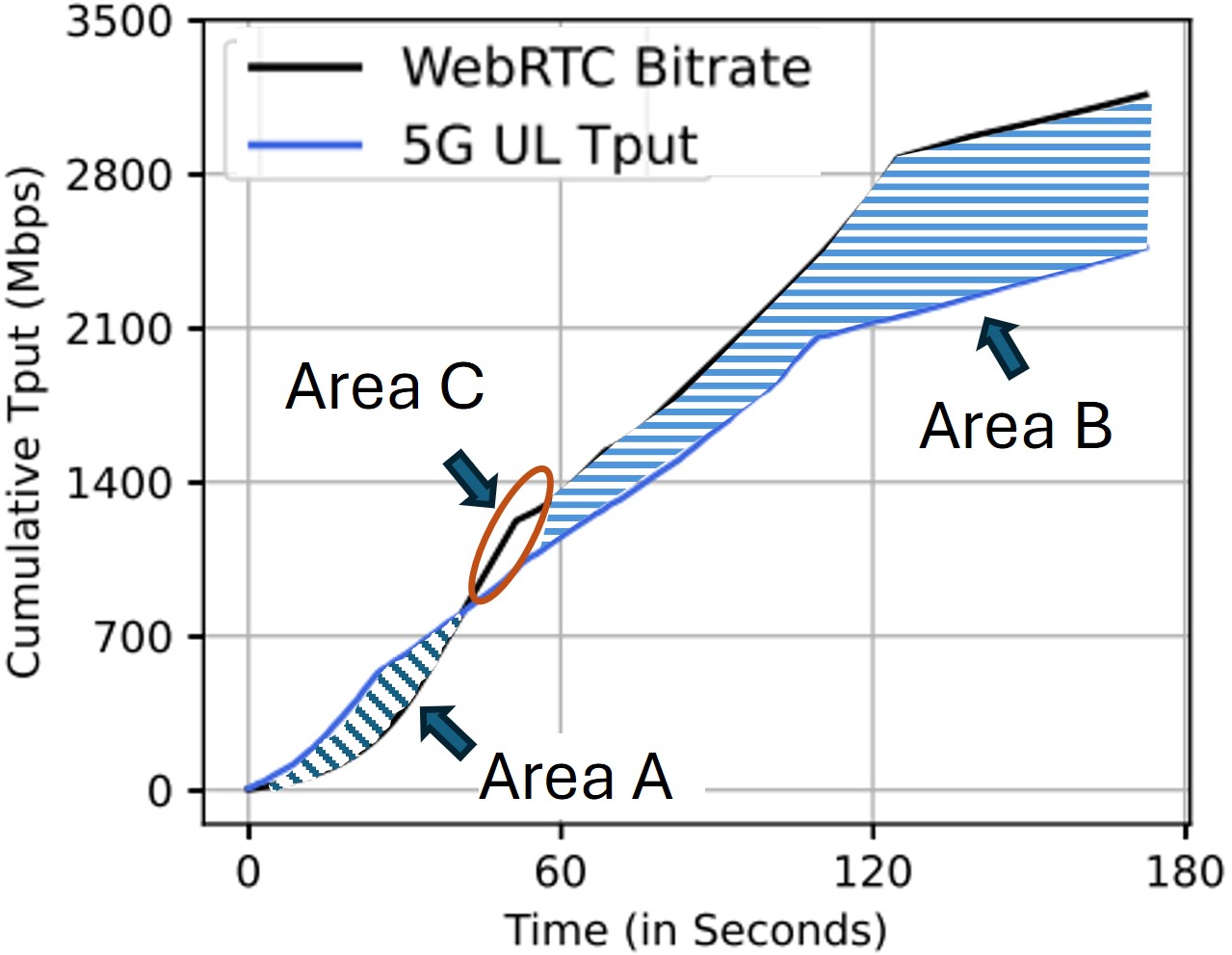}
        \vspace{-2ex}
        \caption{\ac{webrtc}'s Bitrate Choices vs Available Throughput.}
        \label{fig:bitrate_vs_throughput}

        \bigbreak

        \vspace{-2ex}

        \subfloat[RBs Distribution\label{fig:multi-user-rbs}]{{\includegraphics[scale=0.3, keepaspectratio]{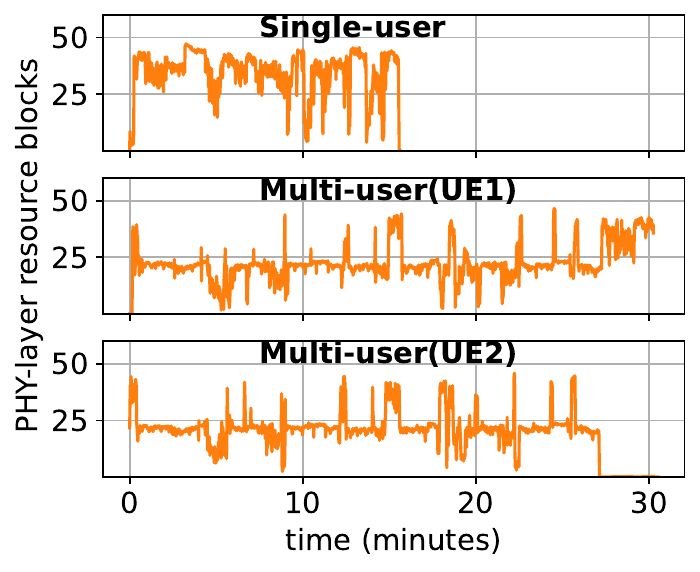}}}
          \hspace{0.1cm}
        \subfloat[Per-Frame Total Delay\label{fig:multi-user-delay}]{{\includegraphics[scale=0.3, keepaspectratio]{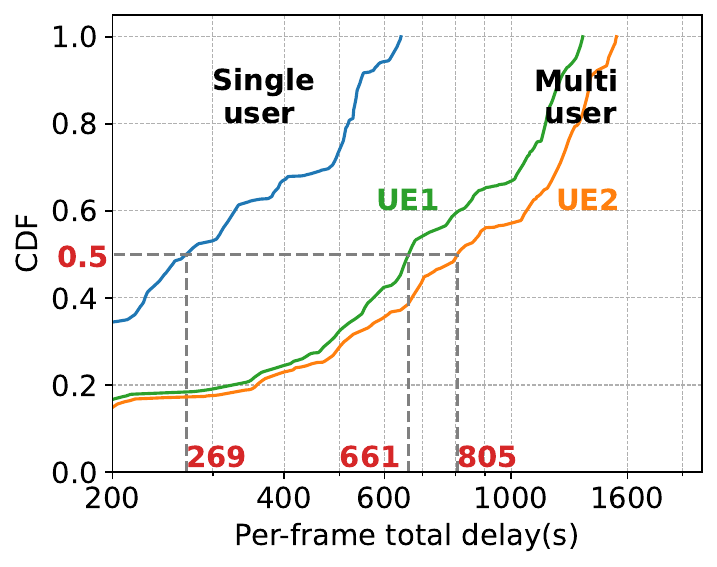}}}%
        \vspace{-2.5ex}
        \caption{Effect of multi-\ac{av}s}
        \label{fig:multi-avs}

    \end{minipage}
   \vspace{-2ex}
\end{figure*}

\section{5G Impact and Network \& Application Adaptation for ToD}
\label{s:cc-adaptation}

Here, we use \ac{webrtc} as a key use case to study how 5G \ac{ran} dynamics impact feedback-based congestion control algorithms for real-time sensor data streaming for \ac{av}s. Specifically, we show that the 5G channel fluctuates at timescales much smaller than the congestion control feedback delay.

\noindent
\simpletitle{PHY Impact on WebRTC Adaptation Mechanisms. }We stream the front-camera to understand how 5G \ac{phy} dynamics influence \ac{webrtc} decision-making. In \fig~\ref{fig:multi-layer-webrtc}, we cross-correlate the \ac{phy} factors with the \ac{webrtc} performance. Plot (i) shows 5G \ac{phy} throughput and \ac{rb} allocation, and plot (ii) shows \ac{webrtc}’s target bitrate. Plots (iii) and (iv) illustrate channel conditions -- as indicated by the \ac{cqi}, \ac{mcs}, and \emph{Per-Frame (\ac{ul}) Network Delay}. 

We see that there is a clear lag (\ie $\sim${10}~seconds) in \ac{webrtc}'s decision -- to explain, in the first 30~seconds, the initial gradual increase in the \ac{phy} throughput is matched by a corresponding graduate increase in the \ac{webrtc}'s target bitrate. Between 30 and 40~seconds, the \ac{phy} throughput suddenly drops within milliseconds as a result of lower \ac{rb}, despite acceptable \ac{cqi} and \ac{mcs} values. This sudden drop causes the video data to queue up in the 5G \ac{ul} channel, leading to a sharp increase in \emph{Per-Frame Network Delay} (0.1–0.22~s) approximately 10 seconds later. At this point, \ac{webrtc} detects congestion and reacts by reducing the target bitrate from 40 to 15 Mbps, which in turn lowers the frame rate. This effect is further amplified by \acp{hos}, both with and without failures, as shown in the inset plots at the bottom of \fig~\ref{fig:multi-layer-webrtc}. Each inset displays the \emph{Per-Frame (\ac{ul}) Network Delay} along with \ac{ho} start and end times. \ac{ho} failures result in retransmissions at the PDCP sub-layer due to packet loss, increasing the \emph{Per-Frame (\ac{ul}) Network Delay} by approximately 20–33\%. Notably, ping-pong \acp{hos} have an even greater impact, as seen in the larger spike in network delay in \fig~\ref{fig:multi-layer-webrtc} plot (iii).

\simpletitle{Need for ``better'' Network Adaptation Mechanism. }To further illustrate the interactions between 5G and \ac{webrtc}'s decision-making process, we use the same trace and plot the cumulative throughput of the \ac{phy} and target bitrate selected by \ac{webrtc} in~\fig~\ref{fig:bitrate_vs_throughput}. Initially, \ac{webrtc} underutilizes the available 5G throughput (area A in~\fig~\ref{fig:bitrate_vs_throughput}), as the target bitrate remains below the network’s capacity, leading to wasted resources. This occurs because \ac{webrtc}'s \ac{gcc} algorithm relies on \ac{rtcp} feedback, which operates on much larger timescales ($\geq$2~seconds by default) than the rapid variations in 5G \ac{phy} throughput (0.5~ms~\cite{rossPam}). As a result, \ac{webrtc} fails to immediately adjust its target bitrate, causing a mismatch that leads to queuing delays and congestion as data accumulates in the buffer.
In area C, as the 5G throughput decreases, \ac{webrtc} continues increasing its target bitrate. Again, due to the delayed feedback, this leads to overutilization (area B), causing spikes in \emph{Per-Frame (\ac{ul}) Network Delay}. These mismatches highlight the inefficiencies in the application layer (\ac{webrtc}) to adapt to 5G’s fast dynamics, underscoring the need for more responsive and precise network adaptation mechanisms. \\

\simpletitle{\textbf{\textit{Implications \& Key Takeaways. }}} The above analysis confirms that application layer-based adaptation and congestion control mechanisms are not sufficient to cope with the ``fast'' 5G channel variability caused by mobility in \ac{av} teleoperations. The 5G \ac{phy} metrics provide another critical dimension to detect problems on the 5G network and react quickly, reducing the \ac{ul} delay. Overall, it is crucial to design applications to be ``5G-aware''.

 %\vspace{-0.5em}

\section{Further Discussion}
\label{s:e2eApproaches}

\noindent
Based on our findings, we conclude our work by discussing the effect of multiple \ac{av}s operating over commercial 5G networks.
The goal is to shed light on possible future directions.

%\vspace{-0.5em}
\subsection{Effect of Multiple AVs}
\label{ss:multi-av}

\noindent
To illustrate the impact of multiple vehicles competing for radio resources, we experiment with two vehicles (users). These results can be generalized to scenarios with more vehicles. To stress the network, both vehicles stream front 64-beam LiDAR data at 277~Mbps. The top plot in \fig~\ref{fig:multi-avs}\subref{fig:multi-user-rbs} shows the number of \ac{phy} \acp{rb} allocated when only one vehicle is streaming. The middle and bottom plots show the \ac{rb} allocation when both vehicles stream simultaneously. As expected, \acp{rb} are nearly halved when two vehicles share the network. This resource competition significantly delays sensor data delivery, as shown in \fig~\ref{fig:multi-avs}\subref{fig:multi-user-delay}. The CDF of \emph{Per-Frame (\ac{ul}) Total Delay} reveals that the median delay more than doubles. Specifically, transmitting the entire LiDAR dataset takes ~15 minutes for a single user but nearly 30 minutes for two. More competing users would further degrade the delay performance.

\vspace{-3ex}
\subsection{Leveraging Multiple Operators}
\label{ss:multi-operators}
When multiple 5G operators are available, we can leverage them to support AV teleoperation. One approach is to utilize MPTCP or MP-QUIC~\cite{ni2023cellfusion} by splitting packets (of each frame) across multiple operators. One issue with this approach is that when one of the operators experiences poor channel conditions, it affects the entire frame, thereby prolonging the \emph{Per-Frame (\ac{ul}) Total Delay}. An alternative approach is to perform operator switching -- namely, utilizing one operator for streaming at any given time, and switching to another operator when the \ac{ul} throughput of the first operator is poor. This strategy can especially help alleviate the poor performance caused by \ac{ho}s, poor \ac{cqi}, or poor coverage of a single operator. 

\begin{figure}[t!]
        \centering
        \includegraphics[scale=0.5, keepaspectratio]{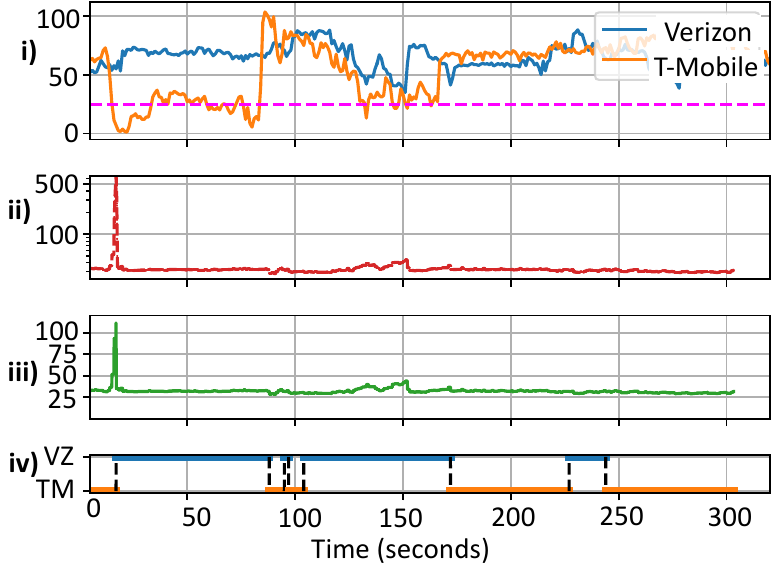}
     %   \vspace{-3ex}% 
        \caption{Multi-Path Streaming for Single Video}
        \label{fig:multi-path}
    \vspace{-3ex}% 
\end{figure}

In Fig.~\ref{fig:multi-path}, we provide an example to
illustrate the performance of operator switching. The top plot (i) shows the \ac{phy} throughput of \ac{tmb} vs. \ac{vz} over time.  The second and third plots show the benefits of operator switching: the second plot
(ii) shows the \emph{Per-Frame Total Delay} and the third plot (iii) shows the \emph{Per-Frame network Delay}. The fourth plot (iv) indicates the operator being used for streaming the current individual frames. We can observe that the \emph{Per-Frame Total Delay} is consistently below 100~ms, except in situations where a sudden throughput drop leads to incorrect bandwidth estimates, making it a promising approach to consider for teleoperation.  

\noindent
\simpletitle{Impact of Data Compression on Downstream AI Tasks.} Sensor data received from AVs may also be used for object detection and tracking to help alert the human teleoperator. We also study the impact of data compression on downstream AI tasks. The details can be found in Appendix \S\ref{aa:ai_tasks}.

%\vspace{-0.5em}
\section{Conclusion}
\label{s:conclude}

We conducted -- to the best of our knowledge -- a first feasibility study of \ac{av} teleoperations over commercial 5G networks in a real-world urban setting, analyzing cross-layer and \ac{etoe} performance. We distinguish our work from previous studies and introduce per-frame level \ac{qoe} metrics to elucidate the impact of key 5G \ac{phy} dynamics -- \ac{cqi}, \acp{bler}, \acp{hos}, and \acp{rb} allocation -- on \ac{etoe} latency and tail performance. Our study reveals the challenges posed by 5G networks and the limitations of existing sensor data streaming mechanisms. While adaptive frame dropping and multi-operator strategies improve tail latency, they cannot fully mitigate poor channel conditions and frequent handovers, especially when driving in a loop and making turns. New 5G features, such as network slicing~\cite{5G-network-slicing}, can provide resource provisioning and prioritized \ac{qoe}, but do not resolve the fundamental limitations of the 5G network. We advocate for a co-designed approach integrating wireless networks, edge/cloud systems, and applications to minimize latency and make \ac{av} teleoperations feasible over live commercial networks. \\

\noindent\textbf{Acknowledgment:} This research was supported in part by NSF under Grants 1915122, 2128489, 2154078, 2212318, 2220286, 2220292, and 2321531, as well as an InterDigital gift.

\bibliographystyle{ACM-Reference-Format}
\bibliography{refs}

% \newpage 
\vspace{-0.15in}
\section{Appendix}
\label{s:append}

\subsection{Ethics}
\label{aa:ethics}
This study was carried out by the research team, volunteers, and paid graduate students. No personally identifiable information (PII) was collected or used, nor were any human subjects involved. Our study complies with the customer agreements of all 5G operators. This work does not raise any ethical issues. 

%\vspace{-0.2in}
\subsection{Experimental Testbed \& Methodology}
\label{aa:method}

\simpletitle{Our \ac{cav} Sensors.  }
We measured the data footprint of several sensors attached to our CAV to establish a baseline for the network support required by a teleoperated vehicle. These footprints are shown in Table \ref{table:raw_throughput}.  We focused on a 128-beam LiDAR, the front-facing FLIR camera, and the combination of front, left, and right cameras for merged video experiments, as these sensors are the primary contributors to teleoperation tasks. Their critical role in perception, coupled with the high volume of data they generate, necessitates significant network throughput. To collect data, we used ROS2 Humble to record both camera and LiDAR outputs, leveraging manufacturer-provided SDKs such as Spinnaker for the cameras and RSlidar SDK for the LiDAR systems.

\begin{table}[t]
    \centering
    \scriptsize
    \centering
\vspace{-2ex}
    \captionof{table}{Our CAV Sensors with Throughput Requirements and Default Sample Rate}
      \vspace{-10pt}
  \begin{tabular}{|c|c|c|c|c|} \specialrule{.15em}{.05em}{.05em}
    \begin{tabular}[c]{c@{}}
     \textbf{Sensor} \\ \textbf{Type}
     \end{tabular}
    & 
    \begin{tabular}[c]{c@{}}
     \textbf{Sensor} \\ \textbf{Model}
     \end{tabular}
    & 
    \begin{tabular}[c]{c@{}}
     \textbf{Raw Data} \\ \textbf{Rates(Mbps)}
     \end{tabular}
    & 
     \begin{tabular}[c]{c@{}}
     \textbf{Raw Data} \\ \textbf{Rates(Frames)}
     \end{tabular} 
    \\ \specialrule{.15em}{.05em}{.05em}
    GPS & Novatel PWRPAK7-E2 GNSS & 0.29 & NA \\ \specialrule{.15em}{.05em}{.05em}
    IMU & 2 OS1-16 and 1 OS1-64 Ousters & 0.038 & 100  \\ \hline
    IMU & Novatel PWRPAK7-E2 GNSS & 0.33  & NA \\ \specialrule{.15em}{.05em}{.05em}
    Odometry & Novatel PWRPAK7-E2 GNSS & 0.28 & NA \\ \specialrule{.15em}{.05em}{.05em}
    Video & Front-FLIR Blackfly S GigE RGB & 24.994 & 30  \\ \hline
    Video & Right-FLIR Blackfly S GigE RGB & 25.467  & 30 \\ \hline
    Video & Left-FLIR Blackfly S GigE RGB & 37.749 * & 30 \\ \hline
    Video & Front-FLIR ADK Thermal & 4.561   & 15 \\ \specialrule{.15em}{.05em}{.05em}
    LiDAR & Robosense Ruby Plus 128 & 125 & 10 \\ \hline
    LiDAR & Right-Ouster OS1-16 & 63.411 & 10  \\ \hline
    LiDAR & Left-Ouster OS1-16 & 64.278 & 10 \\ \hline
    LiDAR & Front-Ouster OS1-64 & 276.814 & 10 \\ \specialrule{.15em}{.05em}{.05em}
    RADAR & Front-Conti ARS 408 & 1.4  & NA \\ \hline
    RADAR & Rear-Conti ARS 408 & 1.71  & NA \\ \specialrule{.15em}{.05em}{.05em}
  \end{tabular}
  \label{table:raw_throughput}
   \vspace{-3ex}
\end{table}

\simpletitle{Testbed}
Our testbed consists of a laptop (representing an \ac{obu}), a USB-tethered smartphone (serving as the 5G radio), and a cloud-hosted server (functioning as our driver station/location). We utilize \ac{webrtc} and two Chrome browser instances, each with Python logging servers co-located on each side, to capture logs. Additionally, we use OBS to stream our CAV camera data as a webcam for use in the \ac{webrtc} client. We record the resulting video on the server side using JavaScript recording libraries. 

\begin{figure}[]
  \centering
  \includegraphics[width=0.43\textwidth]{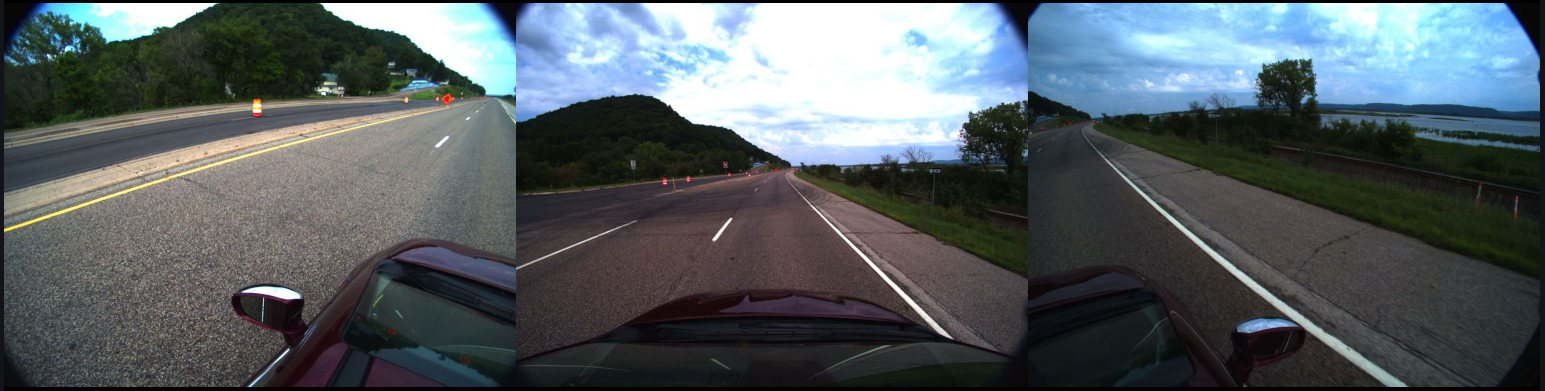}
  \vspace{-1ex}
  \caption{Merged Frames from Front Left, Front, and Front Right Camera.}
  \vspace{-5ex}
  \label{fig:merged_frame}
\end{figure}
% \vspace{-5ex}

\subsection{Teleoperation Command \& Control}
\label{aa:cnc}
In our system, we utilize gRPC, which is an industry standard \ac{cc} framework for teleoperation control. The underlying system utilizes HTTP/2 served over TCP to facilitate command transmission. We record commands from a Logitech simulator platform and replay them over gRPC to mimic a real set of drive commands. This provides a reasonable simulation of drive commands delivered over the network simultaneously with our video streams. This provides a full picture of simulated driving.

\noindent
\begin{figure}[t] 
    \hfill
    \begin{subfigure}[b]{0.24\textwidth} % Adjust width to fit properly
        \centering
        \includegraphics[width=\textwidth, height=1.3in]{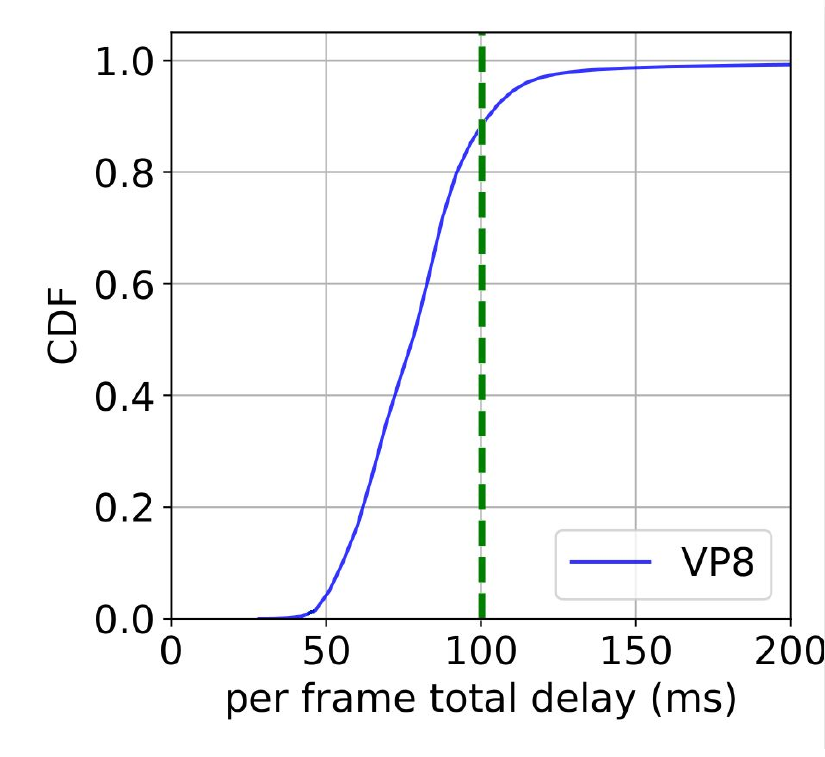}
        \vspace{-4ex}
        \caption{Per-Frame Total Delay}
        \label{fig:merged_video_webrtc_1}
    \end{subfigure}
    \hfill
    \begin{subfigure}[b]{0.23\textwidth}
        \includegraphics[width=\textwidth, height=1.3in]{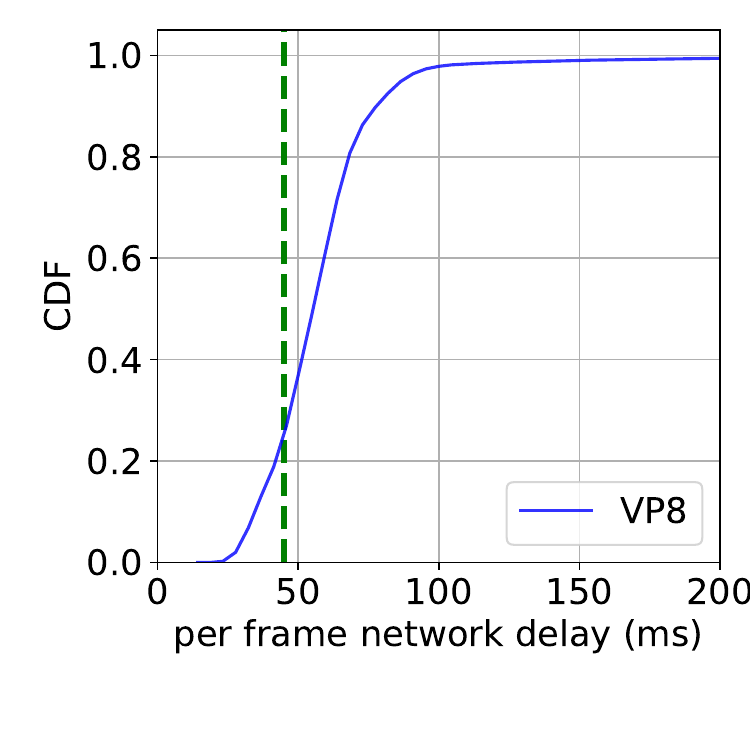}
        \vspace{-4ex}
        \caption{Per-Frame Network Delay}
        \label{fig:merged_video_webrtc_2}
    \end{subfigure}
    \vspace{-5ex}
    \caption{WebRTC QoE for merged-video streaming}
    \label{fig:merged_video_webrtc}
    \vspace{-2ex}
\end{figure}

\begin{figure*}[ht]
  \centering
     \begin{subfigure}{0.15\textwidth}
        \includegraphics[width=\linewidth, height=1.3in]{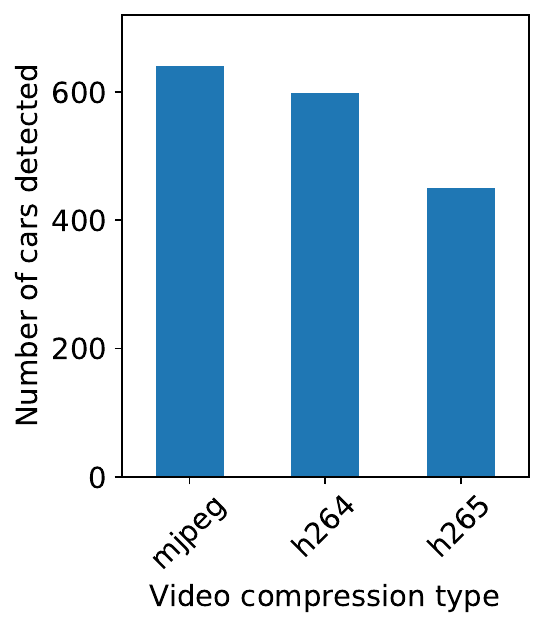}
        \vspace{-2em}
        \caption{\small Object Detection for Cars.}
        \label{fig:detection-counts}
    \end{subfigure}
    \hfill
    \begin{subfigure}{0.22\textwidth}
        \includegraphics[width=\linewidth, height=1.3in]{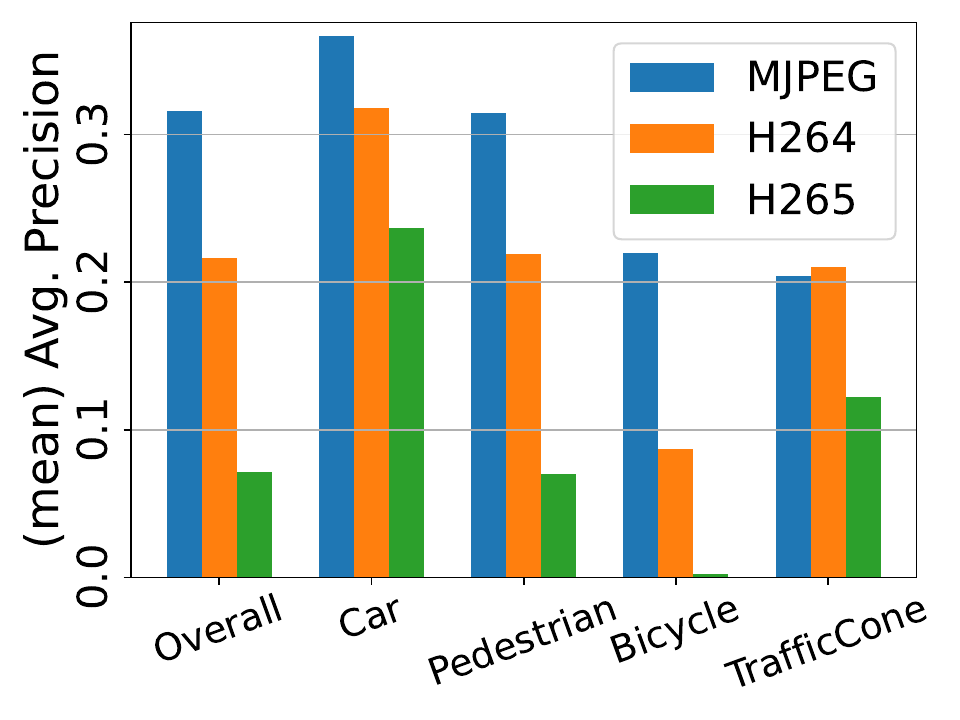}
        \vspace{-2em}
        \caption{Camera-Only Object Detection.}
        \label{fig:obj-det-camera}
    \end{subfigure}
    \hfill
    \begin{subfigure}{0.22\textwidth}
        \includegraphics[width=\linewidth, height=1.3in]{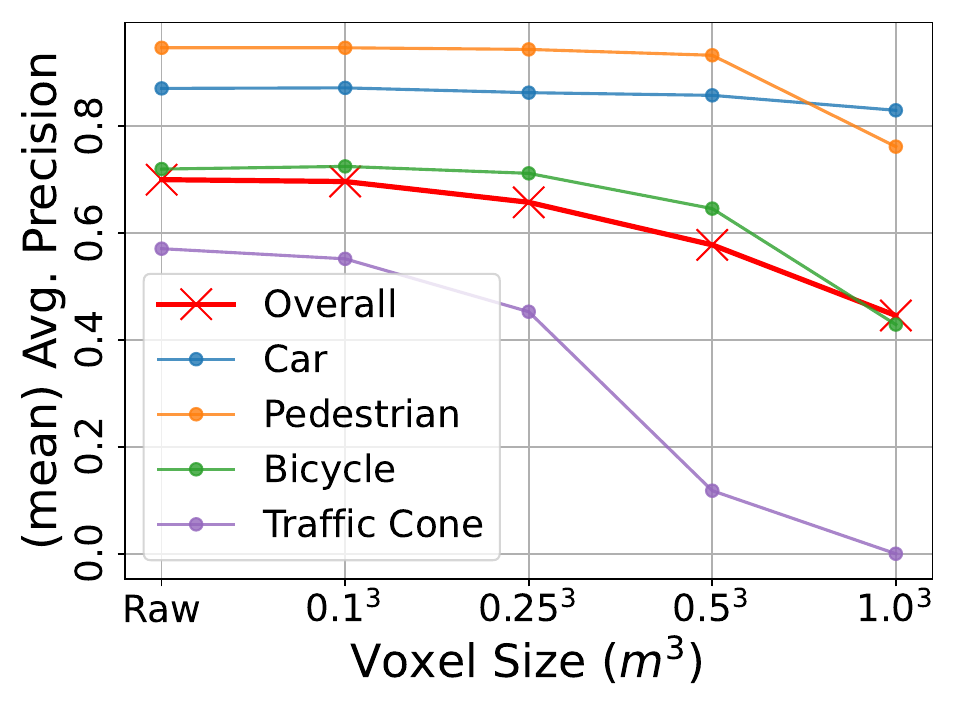}
        \vspace{-2em}
        \caption{LiDAR-Only Object Detection.}
        \label{fig:obj-det-lidar}
    \end{subfigure}
    \hfill
    \begin{subfigure}{0.24\textwidth}
        \includegraphics[width=\linewidth, height=1.3in]{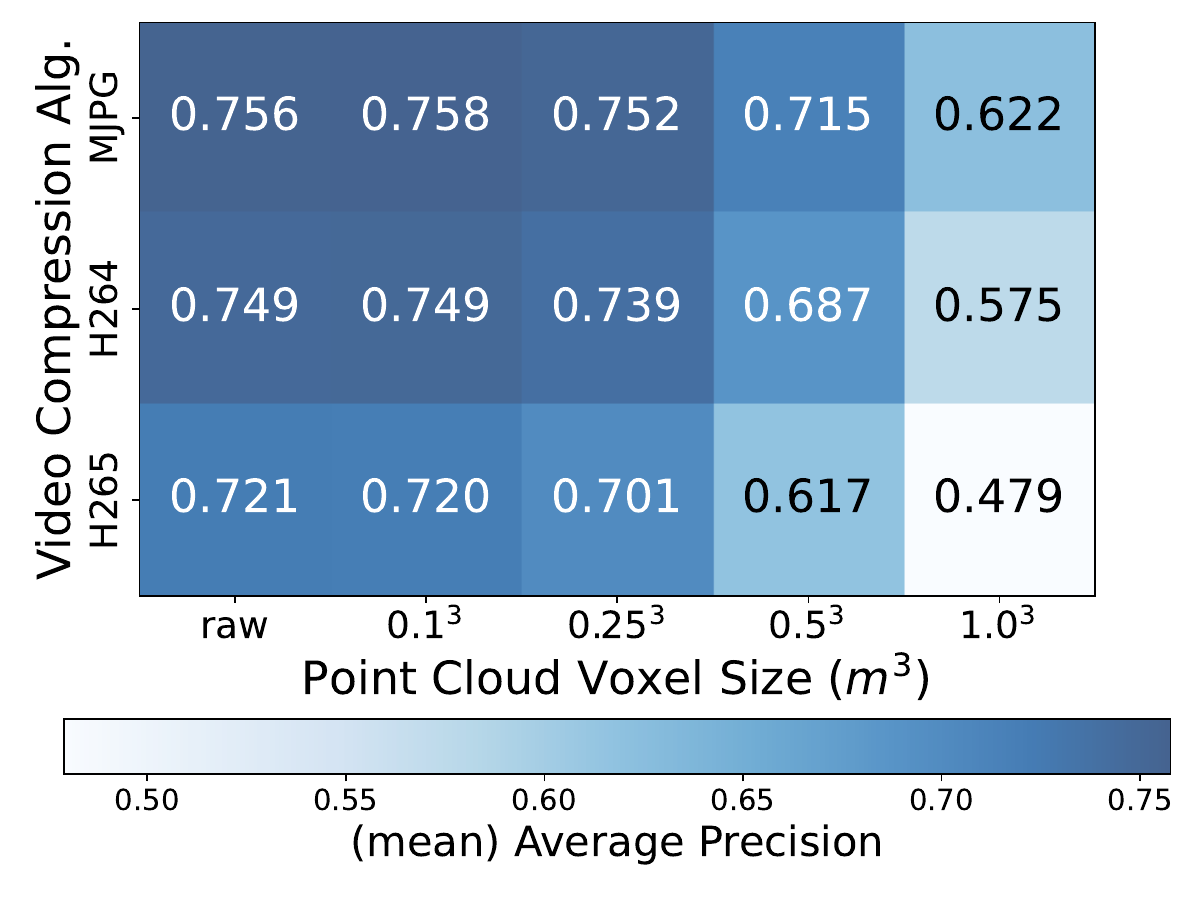}
        \vspace{-2em}
        \caption{Multi-Modality Object Detection.}
        \label{fig:multi-modal-obj-det}
    \end{subfigure}
    \vspace{-3ex}
  \caption{Object Detection with Different Data Modalities and
    Compression Qualities.}
  \label{fig:detection-results}
    % \vspace{-7pt}
\end{figure*}

\subsection{Teleoperation Latency Performance}
\label{aa:perform}
We include in this section some additional results and context for our video streaming latency experimentation. 
Fig.~\ref{fig:merged_frame} illustrates a frame from the merged video stream combining inputs from the front-left, front, and front-right cameras of the \ac{cav} vehicle. This merged video is transmitted over a 5G network to the remote vehicle control station, enabling comprehensive situational awareness for teleoperation. 
The results from our experiments with merged-video streaming using VP8 codec are illustrated in Fig.~\ref{fig:merged_video_webrtc}. We analyze \ac{webrtc} metrics for large-throughput video streaming. Fig.~\ref{fig:merged_video_webrtc}(a) depicts the per-frame total delay which remained below the 100~ms threshold for about 90\% of the frames. Fig.~\ref{fig:merged_video_webrtc}(b) focuses specifically on per-frame network delay, around 20\% of the frames remained below the 45~ms threshold, and around 65\% of frames were in the range of 50-100 ms. The total per-frame delay was slightly higher due to additional processing overhead, yet still largely within acceptable limits for real-time applications.

\begin{figure}[t]
 \vspace{-2ex}
  \centering
  \begin{subfigure}[b]{0.29\linewidth}
    \centering
    \includegraphics[width=\textwidth, height=1.25in]{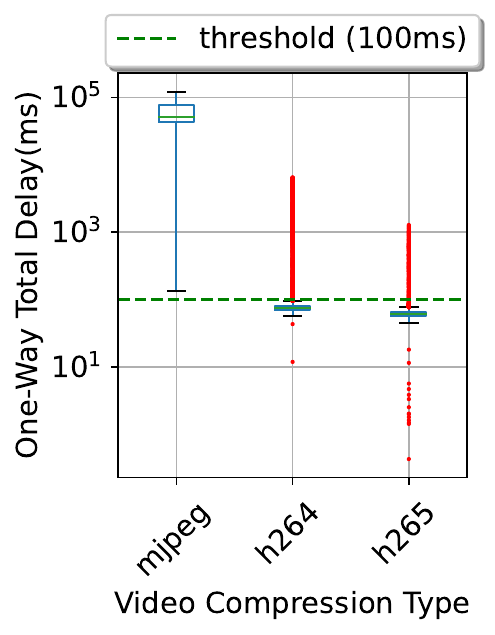}
    \caption{\small Per-Frame Total Delay}
    \label{fig:one_way_app_layer_delay_merged}
  \end{subfigure}
  \hfill
  \begin{subfigure}[b]{0.29\linewidth}
    \centering
    \includegraphics[width=\textwidth, height=1.25in]{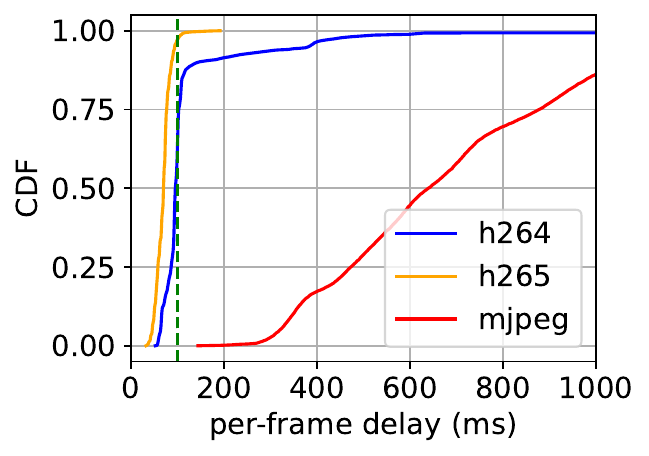}
    \caption{\small Per-Frame Network Delay}
    \label{fig:per_frame_network_delay_merged}
  \end{subfigure}
  \hfill
  \begin{subfigure}[b]{0.33\linewidth}
    \centering
    \includegraphics[width=\textwidth, height=1.25in]{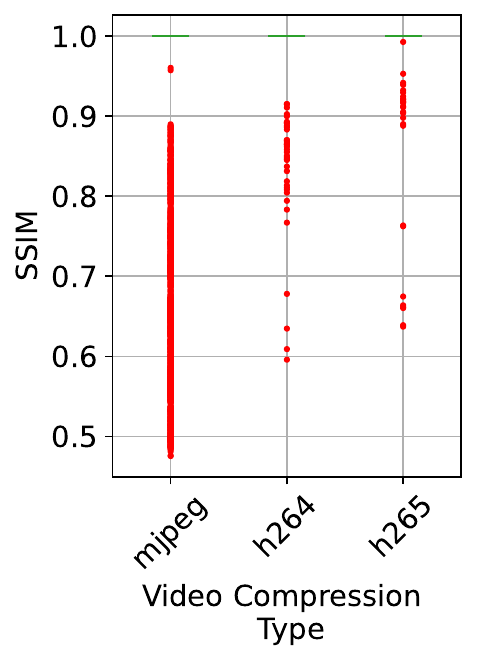}
    \caption{\small Perceptual Quality-Deviation}
    \label{fig:perceptual_quality_deviation_merged}
  \end{subfigure}
   \vspace{-2ex}
  \caption{Merged-videos streaming QoE performance}
  \label{fig:merged_cameras_streams}
  % \vspace{-6ex}
\end{figure}
\noindent
\noindent

Fig.~\ref{fig:merged_cameras_streams} evaluates the \ac{qoe} for merged-video streaming using different video compression methods, including MJPEG, H.264, and H.265. Fig.~\ref{fig:merged_cameras_streams}(a) compares the one-way total delay for each codec, showing that both H.264 and H.265 achieved lower delays compared to MJPEG, with most values below the 100 ms threshold. Fig.~\ref{fig:merged_cameras_streams}(b) analyzes the per-frame network delay for these codecs, where H.265 demonstrates the most consistent performance, while H.264 exhibits slightly higher variability. Lastly, Fig.~\ref{fig:merged_cameras_streams}(c) evaluates the perceptual quality deviation through the Structural Similarity Metric (SSIM), showing that H.264 and H.265 maintain higher and more stable quality compared to MJPEG. These findings indicate that H.264 and H.265 are more suitable for real-time teleoperation streaming due to their balance of low delay and high video quality.

%\vspace{-2em}
\subsection{Impact on Downstream AI Tasks for AV Teleoperations}
\label{aa:ai_tasks}
 
The quality of video not only affects human teleoperator perception, but
also the efficacy of downstream AI tasks such as detecting and recognizing
objects. For example, the detected objects in the video  are often
displayed with a bounding box and a label  to alert the human
teleoperator and assist them in timely decision making. With its 3D
representation and depth information, LiDAR  can enhance the
object detection and recognition accuracy. We now evaluate the impacts of data (video/LiDAR) compression on
the efficacy of objection detection and recognition. We also examine
the benefit of combining both video and LiDAR data in such tasks.

\begin{figure}[t]
    % \hfill
    \begin{subfigure}[b]{0.2\textwidth} % Adjust width to fit properly
        \centering
        \includegraphics[width=\textwidth, height=1.1in]{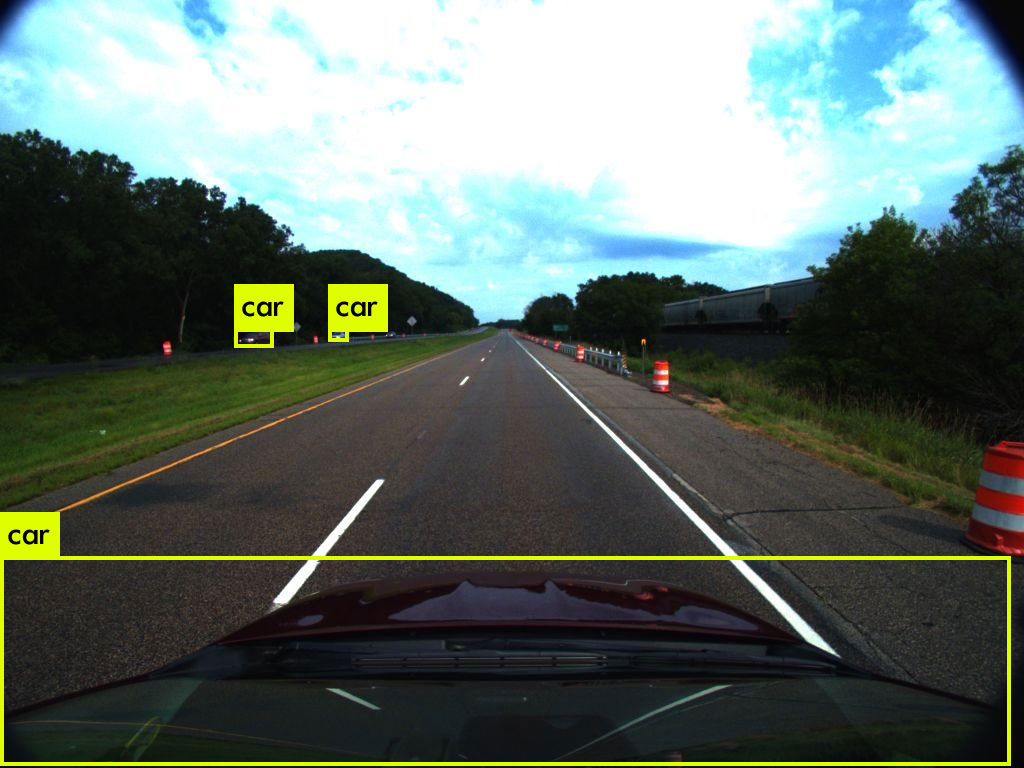}
        \caption{MJPEG Frame Yolo Detection}
        \label{fig:mjpeg_pred}
    \end{subfigure}
    % \hfill
    \begin{subfigure}[b]{0.2\textwidth}
        \includegraphics[width=\textwidth, height=1.1in]{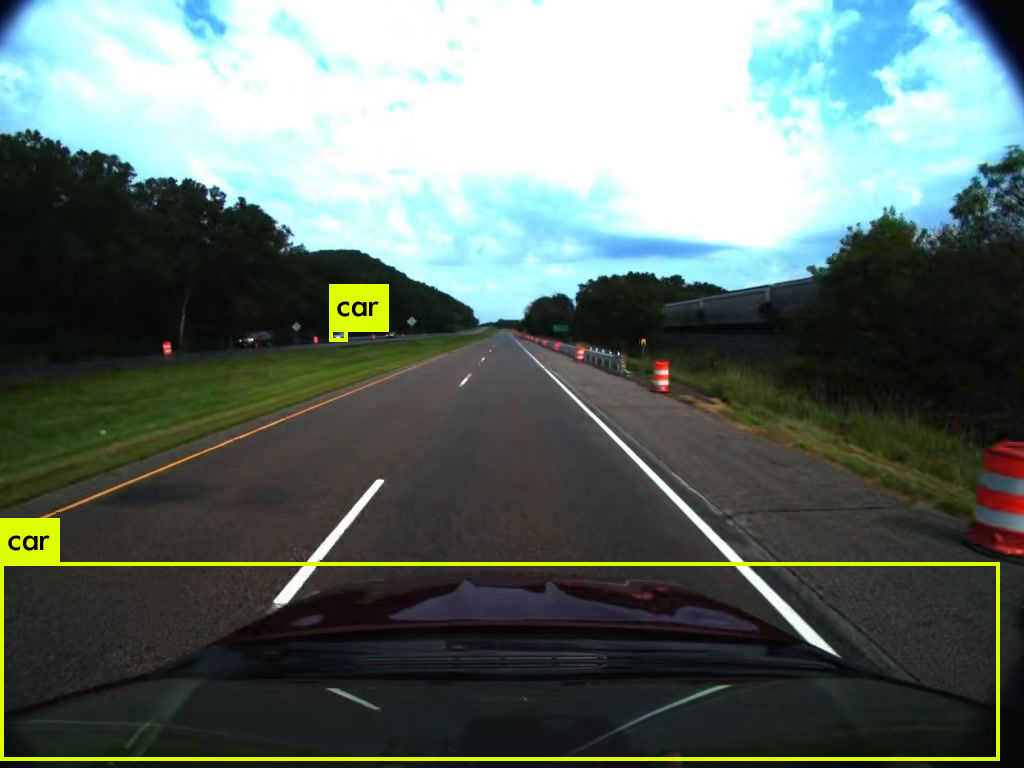}
        \caption{H.265 Frame Yolo \newline Detection}
        \label{fig:h264_pred}
    \end{subfigure}
    \vspace{-2ex}
    \caption{Video Compression Codec Effect on ML Detection }
    \vspace{-0.15in}
    \label{fig:compression_detection}
\end{figure}

To assess the effects of distortion, we
utilize the pre-trained state-of-the-art object detection models: YOLOv8 \cite{YOLO} and 
FocalFormer3D~\cite{chen2023focalformer3d}. As an example,
Fig.~\ref{fig:compression_detection} shows a representative video frame annotated using YOLO:
compared to Fig.~\ref{fig:compression_detection}(a),  one car is not detected
in Fig.~\ref{fig:compression_detection}(b), due to the lower quality of H.265
encoded video. In  Fig.~\ref{fig:detection-results}(a), we compare
the number of objects detected using MJPEG, H.264,
and H.265 video. We see that compared with MJPEG, H.264
and H.265 reduce the number of object detections due to lower video quality. In
particular, while H.265 can significantly reduce the bit rate
requirement (e.g., from using 87.97~Mbps to 4.946~Mbps), this comes with a significant penalty in the efficacy of downstream object detection and recognition.

To quantify this impact, we conduct experiments using the nuScenes
dataset~\cite{nuscenes2019} with ground-truth  labels for various objects
of interest.  Besides  MJPEG, H.264, and H.265 for video compression,
we employ a downsampling technique with  voxel sizes ranging from
$0.1m^3$ to $1m^3$ for LiDAR data. 
Fig.~\ref{fig:detection-results}(b) and Fig.~\ref{fig:detection-results}(c) present
the \emph{mean Average Precision} (mAP) for all classes and four
selected objects types (cars, pedestrians, bicycles, and traffic cones)
using  video-only and LiDAR-only mono-modality 3D object detection,
respectively. Due to the absence of explicit depth information, camera-only
detection performs less effectively than LiDAR, with
0.32 mAP score over all classes, while LiDAR achieves 0.70. 
We can see that lowering the quality of video and LiDAR data  has an
evident effect on 3D object detection and recognition tasks. It is worth noting that such  effect is not
 linear.   For example, H.265 degrades  the overall performance of
 video-based object detection far more aggressively than H.264.
 LiDAR-based detection only experiences a more pronounced decline only after reaching a voxel size of $0.5m^3$. 
 Furthermore, the effects of data compression differ based on object classes. For example, with low H.265 video quality, bicycles become nearly undetectable. Similarly, traffic cones become undetectable in downsampled LiDAR data with a voxel size of $1m^3$.

We further explore the impact of data compression on
\emph{multi-modality} 3D object detection using both video and LiDAR
data. The mAP results for  all classes are shown in
Fig.~\ref{fig:detection-results}(d). 
Multi-modality detection outperforms camera-only and LiDAR-only detection, with 171.0\% and 39.5\% improvements, respectively.
The object detection degradation due to LiDAR compression reduces by 2.8\%-18.6\% when combined with video data. 
In addition to the better object detection performance,
the combination of compressed LiDAR data with 
${0.5m}^3$ voxel size and H.265 video require 25.2\% less
bandwidth than MJPEG encoded video frames alone (65.76~Mpbs vs. 87.97~Mbps). In summary, using compression and bit rate adaptation
may help reduce the per-frame delay (thus increasing the probability
of meeting a target deadline), however their impacts
on efficacy of downstream AI tasks
must be taken into account.  Carefully striking the balance in latency
and video quality is called for.

\end{document}